\documentstyle[epsfig,psfig]{mn}

\newif\ifAMStwofonts

\newcommand{\msyr}{\msun \ {\rm yr^{-1}}}
\newcommand{\msun}{{M}_{\odot}}
\newcommand{\mdotwd}{\dot{M}_{WD}}
\newcommand{\mdot}{\dot{M}}
\newcommand{\mdotscale}{10^{-8} \ \msun \ {\rm yr^{-1}}}

\newcommand{\rmag}{r_{\rm mag}}
\newcommand{\lsun}{{L}_{\odot}}
\newcommand{\sexp}{s_{exp}}

\newbox\grsign \setbox\grsign=\hbox{$>$} \newdimen\grdimen \grdimen=\ht\grsign
\newbox\simlessbox \newbox\simgreatbox
\setbox\simgreatbox=\hbox{\raise.5ex\hbox{$>$}\llap
     {\lower.5ex\hbox{$\sim$}}}\ht1=\grdimen\dp1=0pt
\setbox\simlessbox=\hbox{\raise.5ex\hbox{$<$}\llap
     {\lower.5ex\hbox{$\sim$}}}\ht2=\grdimen\dp2=0pt
\def\simgt{\mathrel{\copy\simgreatbox}}\def\simlt{\mathrel{\copy\simlessbox}}

\ifoldfss
  \ifCUPmtlplainloaded \else
    \NewTextAlphabet{textbfit} {cmbxti10} {}
    \NewTextAlphabet{textbfss} {cmssbx10} {}
    \NewMathAlphabet{mathbfit} {cmbxti10} {} 
    \NewMathAlphabet{mathbfss} {cmssbx10} {} 
  \fi
  \ifAMStwofonts
    \ifCUPmtlplainloaded \else
      \NewSymbolFont{upmath} {eurm10}
      \NewSymbolFont{AMSa} {msam10}
      \NewMathSymbol{\upi}     {0}{upmath}{19}
      \NewMathSymbol{\umu}     {0}{upmath}{16}
      \NewMathSymbol{\upartial}{0}{upmath}{40}
      \NewMathSymbol{\leqslant}{3}{AMSa}{36}
      \NewMathSymbol{\geqslant}{3}{AMSa}{3E}

       \let\le=\leqslant
       \let\ge=\geqslant
    \fi
  \fi
\fi 

\ifnfssone
  \newmathalphabet{\mathit}
  \addtoversion{normal}{\mathit}{cmr}{m}{it}
  \addtoversion{bold}{\mathit}{cmr}{bx}{it}
  \newmathalphabet{\mathbfit} 
  \addtoversion{normal}{\mathbfit}{cmr}{bx}{it}
  \addtoversion{bold}{\mathbfit}{cmr}{bx}{it}
  \newmathalphabet{\mathbfss} 
  \addtoversion{normal}{\mathbfss}{cmss}{bx}{n}
  \addtoversion{bold}{\mathbfss}{cmss}{bx}{n}
  \ifAMStwofonts
    \ifCUPmtlplainloaded \else
      \UseAMStwoboldmath
      \makeatletter
      \new@mathgroup\upmath@group
      \define@mathgroup\mv@normal\upmath@group{eur}{m}{n}
      \define@mathgroup\mv@bold\upmath@group{eur}{b}{n}
      \edef\UPM{\hexnumber\upmath@group}
      \new@mathgroup\amsa@group
      \define@mathgroup\mv@normal\amsa@group{msa}{m}{n}
      \define@mathgroup\mv@bold\amsa@group{msa}{m}{n}
      \edef\AMSa{\hexnumber\amsa@group}
      \makeatother
      \mathchardef\upi="0\UPM19
      \mathchardef\umu="0\UPM16
      \mathchardef\upartial="0\UPM40
      \mathchardef\leqslant="3\AMSa36
      \mathchardef\geqslant="3\AMSa3E

       \let\le=\leqslant
       \let\ge=\geqslant
    \fi
  \fi
\fi 

\ifnfsstwo
  \DeclareMathAlphabet{\mathbfit}{OT1}{cmr}{bx}{it}
  \SetMathAlphabet\mathbfit{bold}{OT1}{cmr}{bx}{it}
  \DeclareMathAlphabet{\mathbfss}{OT1}{cmss}{bx}{n}
  \SetMathAlphabet\mathbfss{bold}{OT1}{cmss}{bx}{n}
  \ifAMStwofonts
    \ifCUPmtlplainloaded \else
      \DeclareSymbolFont{UPM}{U}{eur}{m}{n}
      \SetSymbolFont{UPM}{bold}{U}{eur}{b}{n}
      \DeclareSymbolFont{AMSa}{U}{msa}{m}{n}
      \DeclareMathSymbol{\upi}{0}{UPM}{"19}
      \DeclareMathSymbol{\umu}{0}{UPM}{"16}
      \DeclareMathSymbol{\upartial}{0}{UPM}{"40}
      \DeclareMathSymbol{\leqslant}{3}{AMSa}{"36}
      \DeclareMathSymbol{\geqslant}{3}{AMSa}{"3E}

       \let\le=\leqslant
       \let\ge=\geqslant
    \fi
  \fi
\fi 

\ifCUPmtlplainloaded \else
  \ifAMStwofonts \else 
    \def\upi{\pi}
    \def\umu{\mu}
    \def\upartial{\partial}
  \fi
\fi

\title[Rapid variability in symbiotic binaries]{A search for rapid
photometric variability \\ in symbiotic binaries
}
\author[Sokoloski, Bildsten, \& Ho]
       {J.~L.~Sokoloski,$^1$ Lars Bildsten,$^2$ and Wynn C. G. Ho$^3$\\
        $^1$Department of Physics and Astronomy, Southampton
University, Highfield, Southampton SO17 1BJ, UK\\
	$^2$Institute for Theoretical Physics and Department of Physics, Kohn Hall, University of
California, Santa Barbara, CA 93106 \\
	$^3$Center for Radiophysics and Space Research, 
Department of Astronomy, Cornell University, Ithaca, NY 14853}

\date{Accepted .
      Received;
      in original form }

\pagerange{\pageref{firstpage}--\pageref{lastpage}}
\pubyear{2000}

\begin{document}

\maketitle

\label{firstpage}

\begin{abstract}

We report on our survey for rapid (time scale of minutes) photometric
variability in symbiotic binaries.  These binaries are becoming an
increasingly important place to study accretion onto white dwarfs 
since they are candidate Type Ia supernovae progenitors.
Unlike in most cataclysmic variables, the white dwarfs in symbiotics
typically accrete from a wind, at rates greater than or equal to
$10^{-9} \msyr$.  In order to elucidate the differences between
symbiotics and other white dwarf accretors, as well as search for
magnetism in symbiotic white dwarfs, we have studied 35 symbiotic
binaries via differential optical photometry.  Included in our sample
are all but one of the symbiotics from the lists of \nocite{ken86}
Kenyon (1986) and
\nocite{dk88} Downes \& Keyes (1988)
with published $V$ magnitude less than 14 and declination greater than
$-20^\circ$.  Our study is the most comprehensive to date of rapid
variability in symbiotic binaries.  We have found one magnetic
accretor, Z And, previously reported by \nocite{sb99} Sokoloski \&
Bildsten (1999).  In four systems (EG And, BX Mon, CM Aql, and BF
Cyg), some evidence for flickering at a low level (roughly 10 mmag) is
seen for the first time.  These detections are, however, marginal.
For 25 systems, we place tight upper limits on both aperiodic
variability (flickering) and periodic variability, highlighting a
major difference between symbiotics and cataclysmic variables.  The
remaining five of the objects included in our sample (the 2 recurrent
novae RS Oph and T CrB, plus CH Cyg, $o$ Ceti, and MWC 560) had
previous detections of optical flickering.  We discuss our extensive
observations of these previously-known flickering systems in a
separate paper.  Five new variable stars were discovered
serendipitously in the fields of the survey objects, and the
observations of these stars are also presented elsewhere.  We discuss
the impact of our results on the ``standard'' picture of wind-fed
accretion, and speculate on the possibility that light from
quasi-steady nuclear burning on the surface of the white dwarf hides
the fluctuating emission from accretion.

\end{abstract}

\begin{keywords}
accretion, accretion discs --- methods: data analysis --- surveys --- binaries:
symbiotic --- stars: magnetic fields --- stars: oscillations 
\end{keywords}

\section{Introduction}

Symbiotic stars, or symbiotic systems (SS), are wide binaries in which
material is transferred from an evolved red giant star to either a white
dwarf (WD), main-sequence star, or in a few cases, a neutron star
\nocite{ken86} (Kenyon 1986). The material can be transferred either as the red giant
overflows its Roche lobe, or by Bondi-Hoyle capture of the red-giant
wind (see \nocite{yun95} Yungelson et al. 1995 for modeling and
population synthesis analysis of wind-fed SS).  In most symbiotics,
Bondi-Hoyle capture of the red-giant wind by the WD is the mode of
mass transfer (Iben \& Tutukov 1996; Luthardt
1992)\nocite{it96,lut92}. Radiation from the hot component turns the
red-giant wind into a partially ionized nebula, producing the
characteristic ``symbiotic'' optical spectrum with high ionization
state emission lines superimposed on the cool red-giant continuum
\nocite{ken86} (Kenyon 1986).

Since most SS contain white dwarfs (sometimes in a state similar to
white dwarfs at the centers of planetary nebulae; \nocite{kw84,mur91}
Kenyon \& Webbink 1984; M\"urset et al. 1991), it is important to
place symbiotics within the context of other binaries with accreting
white dwarfs. Selected properties of several types of accreting-WD
systems --- cataclysmic variables (CVs), supersoft X-ray sources, and
symbiotics --- are listed in Table \ref{tab:wdacc}.  It is clear that
the three types of systems form a hierarchy of time scales, sizes, and
accretion rates.  Our survey for rapid variability in symbiotics
represents part of an attempt to understand some of the
phenomenological differences between symbiotics and the other WD
accretors.

The mass-donor stars in CVs are low-mass main sequence stars which
over-flow their Roche lobes. For CVs with orbital periods shorter than
2 hours, the time-averaged accretion rate of $\sim 10^{-10} M_\odot \
{\rm yr^{-1}}$ is set by gravitational radiation losses.  For CVs with
orbital periods in excess of 3 hours, where magnetic winds are
presumably driving angular momentum losses (\nocite{war95}Warner
1995), the time-averaged accretion rate is larger by nearly an order
of magnitude. These CV-accretion rates are too low for
thermally-stable steady burning of the accreting fuel, and so this
fuel is burned explosively in classical novae outbursts. In the
classical supersoft X-ray sources, on the other hand, unstable mass
transfer from an evolved main sequence star with mass in excess of a
solar mass leads to accretion onto the WD at rates set by the
Kelvin-Helmholtz time, $\dot M\sim 10^{-8}-10^{-6} M_\odot \ {\rm
yr^{-1}}$ (\nocite{kvdh97}Kahabkha \& van den Heuvel 1997).  These
rates are high enough for stable burning of the accreting fuel
(\nocite{vdh92}van den Heuvel et al. 1992).  Thus, the long-term
optical variations seen in supersoft sources must be due to something
other than a thermonuclear runaway, such as a change in the size of
the WD photosphere in response to a change in the accretion rate
(e.g. as in the system RX J0513.9-6951;
\nocite{kvdh97}Kahabka \& van den Heuvel 1997).  In symbiotics,
Bondi-Hoyle capture of the red giant wind gives $\mdot \sim
\mdotscale$ onto the white dwarf, which could produce steady
nuclear burning in some systems (\nocite{siosta94}Sion \& Starrfield 1994).  

There are at least 3 kinds of outbursting symbiotics.  Symbiotic slow
novae (or just 'symbiotic novae') have experienced a single,
decades-long thermonuclear event.  Recurrent novae have experienced
multiple events of much shorter duration that are most likely of
thermonuclear origin
\nocite{mk92} (Miko{\l}ajewska \&
Kenyon 1992b).  Most symbiotics, however, show smaller ``classical
symbiotic outbursts'' whose origin is unclear.  These classical
symbiotic outbursts could be related to the quasi-steady burning of
material on the surface of the WD, shell flashes, or unstable
accretion.  Symbiotics can also be divided into several sub-categories
based upon their infrared (IR) colors.  S-type, or ``stellar'' systems
have IR colors like those of isolated field red giants, whereas
D-type, or ``dusty'' systems have redder IR colors indicative of dust.
The IR D-type systems generally contain Mira variables with very high
mass loss rates, and are detectable radio sources \nocite{st90}
(Seaquist \& Taylor 1990).  IR S-types usually have smaller binary
separations than IR D-types, lower rates of mass-loss from the red
giant, and fewer are detected in the radio
\nocite{st90} (Seaquist \& Taylor 1990).  The S-types are more common,
with about 80\% of symbiotics falling into this category.  Most of the
results that will be presented in this paper are for IR S-type systems
that have either classical symbiotic outbursts or no recorded
outbursts.

\begin{table*}
\centering
\begin{minipage}{140mm}
\caption{Symbiotic Binaries in Relation to Other White Dwarf Accretors\label{tab:wdacc}} 
\begin{tabular}{rccc}
\\
\hline
\hline
 & & & \\
 & Cataclysmic Variables & Supersoft Sources & Symbiotics \\
 & & & \\
\hline
 & & & \\
Orbital Period: & Hours & Hours - Days & Years\\
Mass Transfer Mechanism: & Stable RLOF\footnote{RLOF = Roche lobe overflow; TNR = thermonuclear runaway;
DI = disk instability} & Unstable RLOF
& Wind or RLOF\\
$\mdotwd$ ($\msyr$)\footnote{$\mdotwd$ is the time-averaged accretion
rate onto the white dwarf.}:& $ 10^{-10}-10^{-8}$ & $ 10^{-8}-10^{-6}$ &  $ 10^{-9}-10^{-5}$\\
Observed Number: &  400-500 & $\approx 35$ & $\approx 190$ \\
Magnetic Subclass: & Yes  & ? & Yes \\
Outbursts: & TNR$^a$ \& DI$^a$ & Cause? & Cause? \\
Disk: & Yes & Yes & Some? \\
Steady Nuclear Burning: & No & Yes & Some \\
Flickering: & Yes & Some & Some \\
\end{tabular}
\end{minipage}
\end{table*}

\subsection{Flickering as a Diagnostic of Nuclear Burning} \label{sec:flickdiag} 

As mentioned above, the different accretion rates in CVs and supersoft
sources cause the fuel on a CV WD to be burned unstably in a nova
explosion, whereas the fuel in at least some supersoft sources appears
to be burned quasi-steadily \nocite{vdh92} (van den Heuvel et
al. 1992). The mass transfer rates in symbiotic systems bridge the gap
between those in CVs and those in supersoft sources, and the thermal
stability of the nuclear burning on the WDs in SS is still an open
issue.  The question of nuclear burning in classical symbiotics is
important because the outbursts in these systems could be caused by
processes related to the burning, and also because quasi-steady
burning could allow the white dwarf to increase in mass until it
approaches the Chandrasekhar limit and explodes as a Type Ia
supernovae.

One way to address this issue, as well as the question of magnetism in
the accreting WDs in SS, is to carry out rapid optical monitoring. The
flickering\footnote{We refer to any stochastic or aperiodic brightness
variations as ``flickering'', even if the amplitude of these
variations is small.  CV-like stochastic variability of up to tenths
of magnitudes is termed ``large-amplitude'' or ``strong'' flickering, whereas the
smaller variations seen in the symbiotics discussed in section
\ref{sec:newcandflick} will be referred to as ``small-amplitude'' or ``weak'' flickering.}
seen in CVs on short (e.g., minute) time scales is presumed to come
from material being accreted onto a white dwarf
\nocite{bru92,bd93,ymw97,lyu97,zb98,fb98} (Bruch 1992; Bruch \& Duschl 
1993; Yonehara et al. 1997; Lyubarskii 1997; Zamanov \& Bruch 1998; Fritz
\& Bruch 1998). The detailed origin of optical flickering in CVs is
not well understood, although proposed mechanisms include unstable
mass transfer from the mass-donor star leading to flickering on a disk
hot spot, magnetic discharges in an accretion disk, turbulence in an
accretion disk, and boundary layer instabilities leading to unsteady
accretion
\nocite{bru92} (Bruch 1992). 

Flickering-type optical variability has also been reported for some
supersoft sources (\nocite{chcs97,vtrpb98,mhsdbm98}Crampton et
al. 1997; van Teeseling et al. 1998; Meyer-Hofmeister et al. 1998).
Nuclear-burning time scales and thermal time scales associated with
nuclear burning on a WD are generally thought to be too long for
changes associated with this burning to produce rapid optical
flickering.  Thus, the rapid variability in supersoft sources may
instead be due to the re-processing of nuclear-burning light (which is
emitted predominantly in the soft X-ray regime) into the optical by a
disk rim whose height, and therefore reprocessing area/volume, changes
rapidly (\nocite{mhsdbm98}Meyer-Hoffmeister et al. 1998).  Nuclear
burning on the surface of a WD is also a major source of luminosity in
some SS.  Reprocessed into the optical by the surrounding nebula, it
could therefore significantly affect the optical variability
properties of symbiotic systems.

Symbiotics are particularly interesting for the study of rapid
variability because some flicker and some do not.  An early search for
rapid variability in SS was conducted by
\nocite{wal77} Walker (1977), who observed 16 southern symbiotics from the South
African Astronomical Observatory.  Using a 0.5 and a 1 meter
telescope, with typical observations lasting about 30 minutes, and
with a time resolution of 1 - 5 seconds, he found that only the
symbiotic recurrent novae (which by the nature of their thermonuclear
outbursts are clearly not steadily burning fuel\footnote{A possible
exception is T Pyx \nocite{pat98} (Patterson et al. 1998).})  had
rapid variability with amplitude greater than his detection limit of 1
- 2\%.  In 1996, a survey of 8 northern hemisphere symbiotics was
performed by \nocite{dob96} Dobrzycka et al. (1996).  With
observations up to almost 5 hours in length, they detected rapid
variability in one of the two recurrent novae they observed, RS Oph,
plus they detected variability in two additional objects: MWC 560 and
CH Cyg.  They speculated about a possible inverse correlation between
the hot component luminosity and presence of flickering.  Our survey
builds upon this previous work.

\subsection{Periodic Variability from Magnetic Accretion} 

In CVs with a sufficiently large WD magnetic field, the accretion flow
is funneled onto the WD polar caps, where the mostly radial accretion
forms a stand-off accretion shock.  The heated gas behind the shock
produces X-ray emission, and through reprocessing at the stellar
surface, optical light \nocite{pat94} (Patterson 1994). The
magnetospheric radius, $\rmag$, is where the magnetic pressure of the
WDs dipolar field, $B\approx B_{\rm s}(R/r)^3$ (where $B_{\rm s}$ is
the magnetic field at the WD surface, and $r$ is the distance from the
center of the star), is comparable to the ram pressure of the accreting
matter.  Setting these two pressures equal gives
\begin{eqnarray} \nonumber
{\rmag \over R} &  \approx & 10\left({R \over 10^9 \ {\rm cm}}\right)^{5/7}
\left(\frac{B_{\rm s}}{10^6 \ {\rm G}}\right)^{4/7} \\
 & & \hspace{0.3 cm} \times \;
\left(\frac{\mdotscale}{\mdot}\right)^{2/7} \left( \frac{0.6 \ \msun}{M} \right)^{1/7}, 
\end{eqnarray}
where $R$ is the WD radius, and $M$ is the WD mass. Thus,
if a WD has $B_{\rm s} \simgt 10^5
\ {\rm G}$ (for $\mdot \approx \mdotscale$), it will form a
magnetosphere and, since the magnetic dipole is typically not aligned
with the rotation axis, produce emission that is modulated on the WD
spin period. Indeed, DQ-Her-like objects, which contain magnetized WDs
accreting at rates roughly comparable to those in SS, pulsate in the
optical at $\sim 10-100$ mmag levels (Patterson 1994).  The WD spin
period, $P_{\rm s}$, is determined by the amount of angular momentum
the WD has received from the accreting material, or in other words,
how much it has been ``spun up''. The shortest possible
$P_{\rm s}$ for steady magnetic accretion is the Kepler period at the
magnetosphere, $P_{\rm eq}=2\pi (\rmag^3/GM)^{1/2}$, or
\begin{eqnarray} \label{eqn:peq} \nonumber
P_{\rm eq}& \approx &12 \ {\rm min} \left( {R \over  10^9 \
{\rm cm}} \right)^{18/7}  
\left(\frac{B_{\rm s}}{10^6 {\rm G}} \right)^{6/7}\\
 & & \hspace{0.5cm} \times \;\left(
\frac{\mdotscale}{\mdot}\right)^{3/7} \left(0.6 M_\odot \over M\right)^{5/7}.
\end{eqnarray}
If the WD is spinning faster than $P_{\rm eq}$, the disk material at
the magnetospheric radius will be orbiting less rapidly than the
magnetophere is spinning, and thus will tend to slow the WD down
rather than speed it up.  With a minimal accretion torque $N=\dot
M(GMR)^{1/2}$, the time for a WD in a symbiotic system to reach the
equilibrium spin period, $P_{\rm eq}$, is $t_{\rm spin-up}=2\pi I
/NP_{\rm eq}
\approx 4\times 10^5 \ {\rm yr}(10 \ {\rm min}/ P_{\rm
eq})(\mdotscale/\mdot)$, where $I$ is the moment of inertia of the
white dwarf.  This spin-up time is shorter than the red-giant
lifetime, so the WD can reach $P_{\rm eq}$ if it continually
receives angular momentum at this rate.  Also, from equation
(\ref{eqn:peq}) we see that minutes to tens of minutes are the
interesting time scales for the search for magnetic WDs in SS.  Note
that these same time scales are also the important ones to examine for
some WD pulsations.  The nuclei of planetary nebulae have been seen to
pulsate with period of tens of minutes \nocite{cia96} (Ciardullo \&
Bond 1996), and the first pulsating WD in a CV, with periods ranging
from a few to tens of minutes, was recently found by \nocite{vzyl99}
van Zyl et al. (1999).

There are several motivations for a search for magnetic WDs in SS.
First of all, theories of the origin of magnetism in WDs and theories
of binary stellar evolution indicate that some SS should contain WDs
with strong magnetic fields (\nocite{yun95}Yungelson et al. 1995;
\nocite{sfml88}Sion et al. 1988; \nocite{abl81}Angel et al. 1981).  If
magnetic WDs have evolved from the magnetic Ap and Bp stars, then the
fraction of WDs in binaries such as symbiotics, CVs, and supersoft
sources should be higher than the fraction that are magnetic in the
field (which is about 2\%; \nocite{ans99}Anselowitz et al. 1999).  The
different magnetic fractions should exist because the WDs in
interacting binaries come from progenitor populations that have, on
average, higher masses.  These progenitor populations will therefore
contain a higher fraction of magnetic Ap and Bp stars than the
progenitor population for field white dwarfs.  In fact, 5 - 10 percent
of CVs contain white dwarfs with magnetic fields $B_S \simgt 10^5$ G
(\nocite{wickfer00}Wickramasinghe \& Ferrario 2000).  This high
magnetic fraction could, however, be due to selection effects, as some
CVs are found by their optical or X-ray pulsations.  Magnetic
symbiotics have a different set of selection effects, and therefore
provide an interesting testing ground for these theories.

Secondly, the flickering and other properties of several symbiotics
(CH Cyg and MWC 560) have led \nocite{mski90a}
\nocite{mski90b}Miko{\l}ajewski et al. (1990ab),
\nocite{mm88}Miko{\l}ajewski \& Miko{\l}ajewska (1988),
\nocite{tom92} \nocite{mic93}Tomov et al. (1992), and Michalitsianos
et al. (1993) to propose that these systems contain magnetic
propellers.  The detection of a magnetic WD in a SS by another means,
such as discovery of a coherent oscillation, would allow a test of
these propeller models if the new magnetic system were to experience a
change in accretion rate.  

Finally, the brightness oscillation associated with the spin of a
magnetic, accreting WD in a symbiotic can be a useful diagnostic tool
for these complex systems, as has already been seen in the case of Z
And (\nocite{sb99}Sokoloski \& Bildsten 1999).  Also, such systems provide an
opportunity to study magnetically controlled accretion in an
environment that is quite different from CVs.

\subsection{Our Search for Rapid Variability}

The study of rapid variability from SS can help elucidate the
relationship between SS and other accreting WD systems by providing
direct information about the nature of the accretion region in SS, as
well as the surrounding nebula.  Building upon previous work by other
authors (discussed in \S \ref{sec:flickdiag}), we observed 35 northern
symbiotics between 1997 January and 1999 July.  Five of the survey
objects are previously-known flickerers ($o$ Ceti = Mira AB, MWC 560,
CH Cyg, and the recurrent novae T CrB and RS Oph).  In this paper, we
present our observations for 29 of the survey objects that have no
previous detections of rapid variability.  Seven of these 29 objects
were included in the other searches for rapid variability mentioned
above.  Our observations of the well-known flickerers and the
serendipitous discoveries, including a possible new short-period
$\delta$ Scuti star, are discussed in separate papers
\nocite{sbh01,sbcf01} (Sokoloski et al. 2001a,b).  Our results for Z
And, where we discovered the first stable, coherent oscillation ever
seen in a symbiotic, are also presented in a separate paper
(\nocite{sb99}Sokoloski \& Bildsten 1999).  We describe our
observations and the properties of the survey objects in \S
\ref{sec:surobs}, and the production of light curves in \S
\ref{sec:datareduc}.  The timing analysis and a discussion of sources
of error are presented in \S
\ref{sec:timing}. We display and discuss the results from both the
non-flickering systems and the marginal flickering candidates in \S
\ref{sec:sresults}.  Finally, in \S
\ref{sec:disc} we consider the implications of our results for the
properties of the nebulae, accretion structures, and sources of power
in symbiotic stars.

\section[]{The survey}\label{sec:surobs}

\subsection{Observations} 

All of the observations presented here were performed using the
1-meter Nickel telescope at UCO/Lick Observatory on Mt. Hamilton, near
San Jose.  In order to search for rapid variability, we used
time-resolved differential CCD photometry.  This procedure involved
the repeated measurement of the program star flux relative to an
ensemble average of the fluxes of other stars in the field (the
``comparison stars'').  Each observation consisted of a series of
exposures with either the unthinned, 2048 $\times$ 2048 Loral CCD
(``CCD 2''; 15 $\mu$ pixels), or on several occasions (fewer than 10\%
of the total number of observations) the thinned, 1024 $\times$ 1024
SITe CCD (``CCD 5''; 24 $\mu$ pixels).  The field of view for CCD 2 is
$6.3\arcmin \times 6.3\arcmin$, and the field of view for CCD 5 is
$5.0\arcmin \times 5.0\arcmin$.  Exposure times ranged from 4 to 100
seconds, with a dead time between exposures of roughly 15 to 28
seconds for chip read-out and pre-processing, depending upon the
number of rows read out.  A Johnson-Cousins $B$ filter was used for
most observations.  This filter was selected to reduce the amount of
light detected from the red giant, while still allowing for a large
enough detected flux to produce high signal-to-noise ratio (S/N)
measurements in about one minute.  For several extremely bright
objects, a liquid CuSO$_4$ (+1mm UG2) $U$-band filter was used, and if
multiple observations were done for a single object, a $U$-band
observation was sometimes included.  Table \ref{tab:sobslog} is the
complete log of the observations described in this paper.  A typical
observation lasted 3 to 4 hours, and so 1 to 3 observations were
performed per night.

\begin{table*}
\centering
\begin{minipage}{140mm}
\caption[Observation Log]{Observation Log for Lick Symbiotic Survey}\label{tab:sobslog}
\begin{tabular}{lccccccc}
\hline
\hline
 & & & & & & & \\
Binary & Date, U.T. & Obs. Start & Obs. & $t_{exp}$/ & Number & Filter & Detector\\
Name & (m/d/yr) & ($MJD$) & Length (hr) & $\Delta t$\footnote{$t_{exp}$
is the exposure time, and $\Delta t$ is the average time between
exposure starts.} (s) & of Points\footnote{The number of data points
used in 
 the analysis is equal to the number of images taken minus those
affected by cosmic rays, extremely poor weather, or other 
 problems.  If two values are given, the first is for the light curve, and the second is for the power
 spectrum.} & & (CCD 2 or 5)\\
 & & & & & & & \\
\hline 
 & & & & & & & \\
EG And & 1/17/97 & 50465.150& 0.9 &90.0/109.452 & 31 & $U$& 2\\*
       & 7/9/97 & 50638.450 & 1.2 & 45.0/67.0 & 63 & $U$& 2\\*
       & 7/11/97 & 50640.437 & 1.6 & 4.0/28.0 & 208 & $B$ & 2\\*
       & 9/16/98 & 51072.296 & 5.6 & 90.0/112.855 & 159 & $U$& 5\\
AX Per & 1/23/98 & 50836.127 & 3.9 & 90.0/118.0 & 96 & $B$ & 2\\*
       & 9/15/98 & 51071.354 & 4.5 & 40.0/66.226 & 225 & $B$ & 5\\
V471 Per & 8/22/98 & 51047.430 & 2.3 & 80.0/106.221 & 76 & $B$ & 5\\
S 32 & 3/13/97 & 50520.082$\footnote{The times between data points are only
 approximate on 1997 March 13 due to a clock problem.}$ & 1.3$^b$ & 70.0/$\approx 94.3^b$ & 46 &$B$& 2\\*
     & 9/14/98 & 51070.444 & 2.2 & 70.0/96.191 & 73 & $B$ & 5 \\
UV Aur & 11/2/97 & 50754.447 & 3.2 & 30.0/53.0 & 212 & $B$ & 2 \\
BX Mon & 1/18/97& 50466.288 & 0.3 & 90.0/115.444 &10 &$B$& 2 \\*
       & 2/16/97& 50495.174 & 5.1 & 60.0/85.261 & 208 & $B$ & 2 \\*
       & 3/12/97& 50519.232 & 3.1 & 100.0/125.478 & 87 & $U$& 2 \\*
       & 4/7/97& 50545.165 & 1.0 & 60.0/85.122 & 42 & $B$ & 2 \\*
       & 11/1/97 & 50753.460 & 2.6 & 80.0/102.0 & 88 & $B$ & 2 \\*
TX CVn & 3/13/97 & 50520.180$^b$ & 3.2$^b$ & 60.0/$\approx 87.8 ^b$ & 105 & $B$ & 2 \\*
RW Hya & 2/18/97 & 50497.439 & 2.1 & 30.0/55.478 & 132 & $B$ & 2 \\*
BD-21.3873 & 5/14/97 & 50582.232 & 2.0 & 100.0/114.841 &61 & $B$ & 2 \\*
            & 5/31/98 & 50964.263 & 1.8 & 70.0/90.0 & 68 & $B$ & 2 \\
AG Dra & 6/7/97 & 50606.294 & 3.3 & 22.0/47.212 & 255 & $B$ & 2 \\*
        & 7/23/98 & 51017.203 & 4.7 & 55.0/82.0 & 200 & $B$ & 2 \\
HD 154791  & 8/4/97 & 50664.178 & 2.0 & 7.0/34.0 & 514 & $B$ & 2 \\*
Hen 1341 & 6/28/98 & 50992.222 & 2.3 & 60.0/94.0 & 88 & $B$ & 2 \\
AS 289 & 5/31/98 & 50964.373& 2.6 & 90.0/111.0 & 75 &$B$ & 2 \\*
YY Her & 5/14/97 & 50582.339 & 3.9 & 50.0/75.571 & 168 & $B$ & 2 \\
AS 296  & 8/21/98 & 51046.193 & 3.1 & 90.0/115.010 & 94 & $B$ & 5 \\
S 149 & 5/15/97 & 50583.408 & 2.2 & 60.0/85.404 & 94& $B$& 2 \\*
      & 8/20/98 & 51045.176 & 4.6 & 50.0/76.995 & 205 & $B$ & 2 \\
V443 Her & 4/7/97 & 50545.429 & 2.8 & 60.0/78.433 & 125 & $B$ & 2 \\
FN Sgr & 6/30/98 & 50994.347 & 2.8 & 20.0/48.0 & 210 &$B$& 2 \\ 
CM Aql & 9/16/98 & 51072.153 & 2.9 & 100.0/120.209 & 86 & $B$ & 5 \\
V919 Sgr & 8/22/98 & 51047.206 & 2.5 & 90.0/116.263 & 77 & $B$ & 5 \\
BF Cyg & 4/6/97 & 50544.429 & 2.5 & 50.0/75.317 & 115 & $B$ & 2 \\*
       & 7/11/97 & 50640.212 & 4.1 & 50.0/78.0 & 185 & $B$ & 2 \\*
       & 7/1/98 & 50995.260 & 5.6 & 80.0/108.0 & 176 & $B$& 2 \\*
CI Cyg  & 6/8/97 & 20607.256 & 3.1 & 25.0/43.557 & 252 & $B$ & 2 \\*
        & 9/15/98 & 51071.151 & 4.4 & 100.0/126.2 & 98 & $U$& 5\\
AS 360  & 6/29/98 & 50993.214 & 2.3 & 50.0/78.0 & 104 & $B$ & 2 \\
PU Vul  & 6/15/97 & 50614.442 & 1.4 & 40.0/58.643 & 77 & $U$& 2 \\
He2-467  & 8/22/98 & 51047.326 & 2.1 & 90.0/116.154 & 64 & $B$ & 5 \\
S 180  & 7/23/98 & 51017.434 & 1.5 & 100.0/127.0 & 38 & $B$ & 2\\
S 190  & 8/21/98 & 51046.344 & 3.4 & 50.0/65.995 & 186 & $B$ & 5 \\
AG Peg & 9/17/98 & 51073.148 & 4.5 & 20.0/47.0 & 280,170 & $B$ & 5 \\
R Aqr  & 9/14/98 & 51070.261 & 3.9 & 30.0/51.451 & 265 & $B$ & 5 \\
 & & & & & & & \\
\hline
\end{tabular} 
\end{minipage}
\end{table*}

The exposure times were chosen to maximize the resulting S/N, while
not saturating the program star or any comparison stars, and while
keeping the time between individual exposures to less than 2 minutes.
Longer integrations would have reduced our sensitivity to oscillations
with periods in the interesting range of a few to tens of minutes, and
to the most rapid flickering.  We also generally avoided exposures of
just a few seconds, since for short exposures the comparatively long
read-out time would have made the duty cycle low, and the observations
rather inefficient.

In order to obtain evenly spaced exposures and thus produce data sets
compatible with standard fast Fourier transform (FFT) routines, we
usually employed a timing system developed especially for this project
by Will Deitch. Therefore, for most of the observations, the time
between exposure starts was constant to very high precision.  This
timing system was not available when our earliest observations were
done.  For some of these early observations, the time between
exposures varied by 1 - 2 seconds, or about 5 - 10\%\footnote{The
early version of our timing system introduced some timing error that
affected our sensitivity to periodic oscillations.  Since we arrived
at our pulse amplitude upper limits by repeatedly injecting
oscillations into our real data using the same sampling as the real
light curve, however, the upper limits derived in this way remain
valid.  Comparisons between simulations with even sampling and
slightly uneven sampling, as we had in our early data, indicate that
the sensitivity to periodic oscillations is reduced by a few percent.
}.

\subsection{Differential CCD Photometry}

\begin{figure}
\begin{center}
\epsfig{file=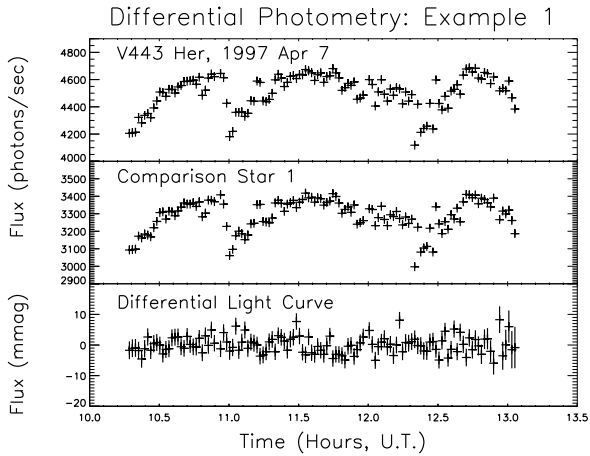,width=8cm}
\end{center}
\caption[Differential Photometry Example 1]{Example of the capabilities of differential
photometry.  In the top panel, the raw light curve for V443 Her from 1997
April 7 (U.T.) is shown.  The raw light curve for a second star in the field is
shown in the middle panel.  In both cases, dips due to clouds or other
atmospheric changes are evident.  In the bottom panel, the differential
light curve formed from the ratio of V443 Her and an ensemble average
of 4 comparison stars is shown.  The atmospheric variability is
completely removed. \label{fig:diffphot1}} 
\end{figure}

\begin{figure}
\begin{center}
\epsfig{file=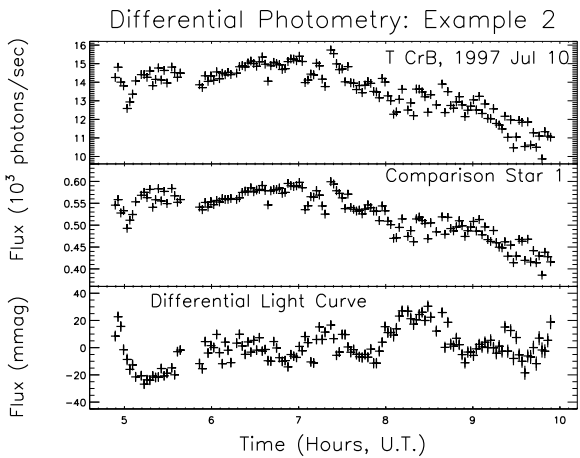,width=8cm}
\caption[Differential Photometry Example 2]{A second example of the capabilities of
differential photometry.  For this observation of T CrB, we were able to
remove the atmospheric effects and see the variability intrinsic to the
source.  The differential light curve was created with an ensemble
average of 4 comparison stars. \label{fig:diffphot2}}
\end{center}
\end{figure}

Differential CCD photometry can yield useful data even when the weather
conditions are not ideal.  Assuming the point-spread function is the
same for each star on the chip, and that clouds reduce the count rate
for each object by the same fraction, the ratio of fluxes extracted from
the same aperture for constant brightness stars should remain
constant despite variable clouds and/or seeing changes.  Two examples of
the usefulness and capabilities of differential photometry are shown in
Figures \ref{fig:diffphot1} and \ref{fig:diffphot2}. 
In Figure \ref{fig:diffphot1}, raw light curves are shown for the
symbiotic V443 Her, another star in the same field of view, and
the differential light curve.  Even though each of the raw
light curves has dips due to clouds or changes in seeing, the differential
light curve is flat to within the error bars.  Using differential
photometry, we can thus place an upper limit on the intrinsic
variability of V443 Her.  In Figure \ref{fig:diffphot2}, raw light
curves for the symbiotic star T CrB and another star in its
field are shown, as well as the differential light curve.  Again, the
raw light curves show variability due to clouds, seeing changes, and
variable air mass.  But this time the ratio shows that T CrB is in fact
intrinsically variable.

\subsection{Sensitivity to an Oscillatory Signal}

Before describing our techniques in more detail, we give an estimate
of our sensitivity.  A typical symbiotic star has a $B\approx 12$. The
photon count rate for an object of magnitude m is given by
\begin{equation}
N_c=N_0  10^{-m/2.5} \times {\rm area} \times
{\rm bandwidth} \times QE,
\end{equation}
where $N_0\approx 1400$ cm$^{-2}$ s$^{-1}$ \AA $^{-1}$ (the flux
density for a $B=0$ object), ``area'' is the mirror collecting area,
``bandwidth'' is the effective bandwidth obtained by integrating over
the transmission of the filter and the atmosphere, and $QE$ is the
quantum efficiency of the CCD.  Plugging in an area of 7850 cm$^2$
(not taking into account the secondary mirror blockage), an effective
bandwidth of 380 \AA (for an airmass of 1.5 and using extinction
tables from
\nocite{hl75} Hayes \& Latham 1975), and a quantum efficiency of 0.15
(rough estimate for Lick's CCD 2), we find $N_c\approx 9900$
photons/sec for $B=12$.  If uncertainties are set purely by photon
counting statistics of the source (e.g., if all comparison stars are
much brighter than the program star, the background is low, and
systematic uncertainties are negligible), then the S/N for detection
of a pulsed signal is given by
\begin{equation} \label{eqn:pulsesn}
\frac{\rm Signal}{\rm Noise} = \frac{(1/4) p N_c t}{\sqrt{N_ct}},
\end{equation}
where $p$ is the pulse fraction, and $t$ is the total integration
time.
The factor of 1/4 in the numerator of equation (\ref{eqn:pulsesn})
comes from the fact that we want to divide the signal counts into at
least 4 phase bins.  If we demand a S/N of 5 (in each phase bin), and
we assume 3 hours of total integration time, then we are sensitive to
pulse fractions of the order
\begin{equation}
p=0.002 \left(\frac{N_c}{9900\, {\rm s}^{-1}}\right)^{-1/2}
\left(\frac{t}{3\, {\rm hr}}\right)^{-1/2},
\end{equation}
for a $B=12$ star. 
This rough estimate is comparable to the result of the more careful
calculation given in equation (\ref{eqn:sens}).  In practice, we cannot
quite reach this level, as explained in the detailed discussion in
\S \ref{sec:datareduc}.    

\subsection{Source Selection}

Table 3 lists the binaries included in our survey.  They were taken
from the lists of
\nocite{ken86} Kenyon (1986) and \nocite{dk88} Downes \& Keyes (1988), 
and were selected primarily because of their observational suitability
for differential photometry timing analysis from Lick Observatory.  In
order to detect mmag level variability, we required roughly $10^5$
stellar counts per integration.  Since we wanted to probe minute-long
time scales, the program star had to provide at least $\sim 800$ c/s
within the extraction region.  This count rate could generally be
achieved for stars brighter than $B \approx 14$ mag with CCD 2, and
for slightly fainter objects with CCD 5 (if the comparison stars were
significantly brighter and therefore did not add much uncertainty).
Note, however, that if there are only a few faint comparison stars
(fainter than the program star at $B$), then these stars and not the
program star set the variability-amplitude detection limit.  Since
typical values of $B-V$ for symbiotics lie between 0 and 1 (see Table
3), and $V$ magnitudes were more readily available, in practice we
selected objects with published $V$ magnitudes of less than 14.

\onecolumn
\begin{table*} \label{tab:surveyprops}
\vspace{2cm}
\epsfig{file=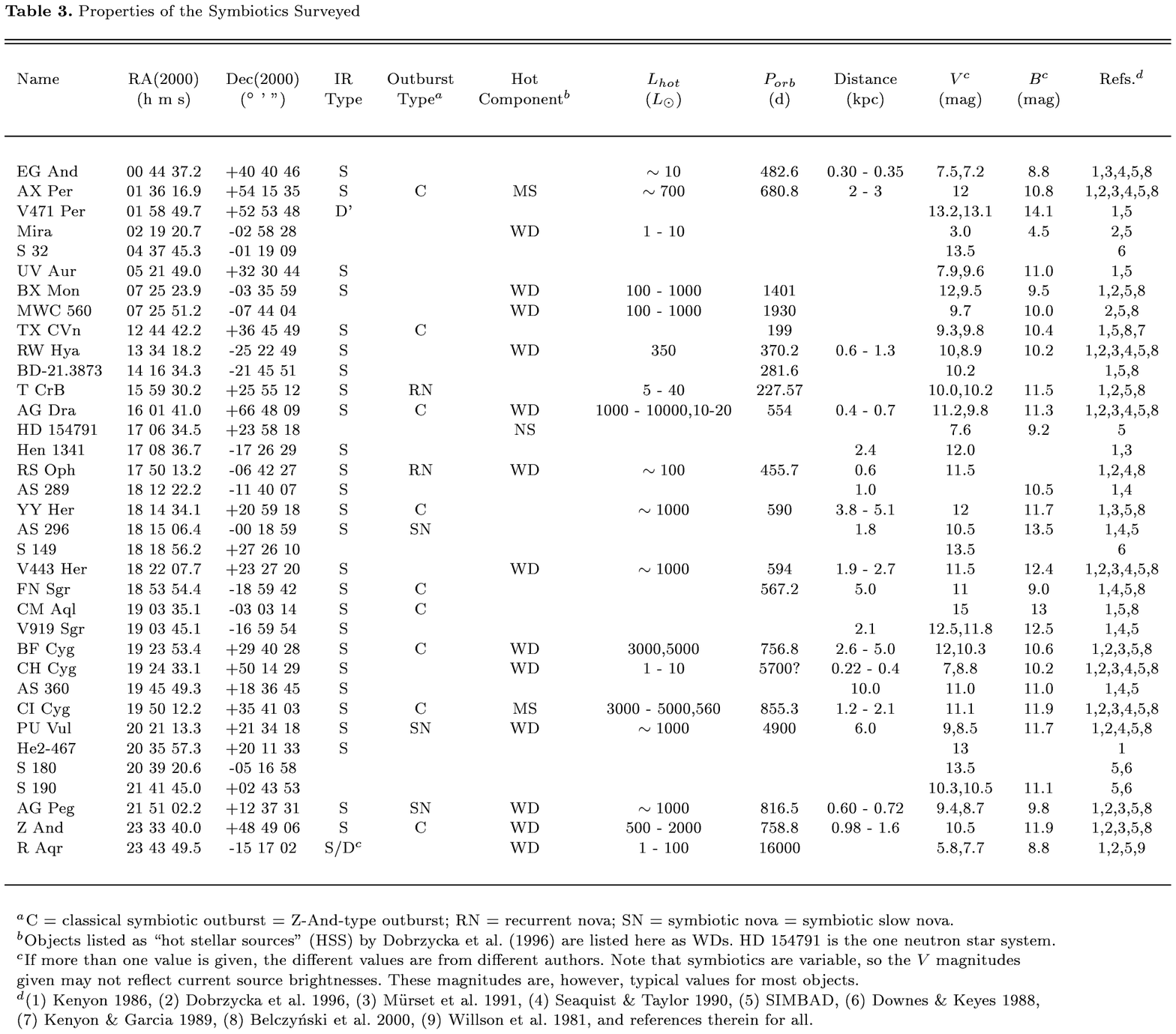,angle=90,width=17cm}
\end{table*}
\clearpage
\twocolumn

For the survey, slow novae still in outburst were generally avoided,
since the expanded envelopes in these systems would hide any rapidly
variable processes near the surface of the WD.  However, a few
slow-novae-like systems and slow novae well after decline from maximum
were observed (PU Vul, AG Peg, and AS 296).  Finally, since Lick
Observatory is at 37$^\circ$ latitude, we could only produce
reasonable light curves for symbiotics with $\delta \simgt -25^\circ$.
In the end, we observed all of the objects from \nocite{ken86} Kenyon
(1986) and \nocite{dk88} Downes \& Keyes (1988) that were not slow
novae, and that have $V < 14$ mag and $\delta > -20$, except one.  In
addition, we observed 2 objects that were slightly more southern (RW
Hya and BD-21.3873), one slightly fainter object (CM Aql), the
symbiotic-like systems Mira AB ($o$ Ceti) and MWC 560, and the
neutron-star symbiotic HD 154791, for a total of 35 objects.  Most of
the symbiotics we observed presumably contain WD accretors
\nocite{mur91} (M\"urset et al. 1991), although the 2 strongest candidate main-sequence systems
(CI Cyg and AX Per; see \nocite{kommsga91} Kenyon et al. 1991,
\nocite{mk92axper}Miko{\l}ajewska \& Kenyon 1992a, and references
therein) are also included in our sample.  All of the symbiotics in
the sample are IR S-types, except for one D$^\prime$-type
system\footnote{The notation D$^\prime$ is sometimes used to indicate
a D-type symbiotic with a yellow absorption line spectrum, as opposed
to a Mira-containing D-type with deep CO and H$_2$0 absorption
bands. (S. Kenyon, private communication).} (V471 Per).  Some
properties of the survey objects are listed in Table 3.  The
previously-known flickerers and Z And are included for comparison,
even though we present our observations of these systems
elsewhere. \nocite{kg89} \nocite{wgm81}

\section{Data reduction}\label{sec:datareduc}

A typical observation, or series of exposures, produced between 50 and
250 images, and the complete survey consisted of over 80 such
observations. Roughly half of these observations were of the
previously-known flickering systems, and the other half are presented
here.  The data reduction and analysis was done in IDL, and it
included the use of software based upon standard IRAF routines.  For
each observation, the data were reduced using the standard techniques
\nocite{gil92} (e.g. Gilliland 1992).

\subsection{Estimation of Uncertainties}

Before describing the aperture photometry and generation of light
curves, we discuss our method for estimating the uncertainties in the
CCD photometry. An understanding of the uncertainties is crucial for
identification of low-level variability intrinsic to the source,
especially low-level flickering-type variability.  We found the formal
uncertainties for each star in each image from the CCD equation
(equation \ref{eqn:ccd}) and Young's expression for expected
atmospheric scintillation.  The uncertainty in CCD aperture photometry
is given by 
\begin{equation} \label{eqn:ccd}
\sigma_{CCD}^2 = c + n_{bins}(1+\frac{n_{bins}}{n_{sky}})(N_S + N_R^2 +
N_D)
\end{equation}
\nocite{how92} (Howell 1992),
where $c$ is the number of integrated stellar source counts in
electrons (or photons), $n_{bins}$ is the area of the software
aperture in bins, $n_{sky}$ is the area of the region used for
background estimation in bins, $N_S$ is the sky (background) counts
(electrons/bin), $N_R$ is the read noise (in rms electrons/bin), $N_D$
is the dark current (electrons/bin).  We refer to bins instead of
pixels, because we always used CCD 2 with $4\times 4$ pixel binning
and CCD 5 with $2\times 2$ pixel binning.  Not included in equation
(\ref{eqn:ccd}) is digitization error, which is less than the value of
the gain (typically 6.8 $e-/ADU$ and 7.7 $e-/ADU$ for our two CCDs),
and therefore always negligible for us.

For our observations, $N_S$ generally ranged from 50 to 500,
read noise squared had the values $N_R^2\approx 100-150$, and the dark
current was $N_D < 1(t_{int}/100 \, {\rm s})$.  Therefore, we can drop
the dark-current term and rewrite the uncertainty as
\begin{equation} \label{eqn:sigma}
\sigma_{CCD}^2 \approx c + n_{bins}(N_S + N_R^2) +
\frac{n_{bins}^2}{n_{sky}} (N_S + N_R^2).
\end{equation}
The number of stellar counts, $c$, was typically a few times $10^4$ to
$10^6$.  The second term, $n_{bins}(N_S + N_R^2)$, which arises from
Poisson fluctuations of the background and readnoise within the
aperture, was approximately $10^4$ for low-background periods, but
could approach $10^5$ when a bright moon and clouds were present.  The
average background count rate for an observation was primarily a
function of the phase of the moon, whether or not there were low
clouds blocking the light from San Jose, and the cloud-cover overhead.
The third term arises from the Poisson uncertainty in the estimation
of the background level, and was a factor of $n_{bins}/n_{sky} \approx
0.1$ smaller than the second term.  In the data analysis, the formal
uncertainties were calculated by replacing the expression $N_S +
N_R^2$ in equation (\ref{eqn:sigma}) with the measured variance from
the sky annulus.  This was a conservative approximation, as the
measured variance should be greater than or equal to $N_S + N_R^2$.

In addition to Poisson uncertainties, there was also an inescapable
contribution to source variability from atmospheric scintillation.
Scintillation refers to the brightness variations of a star viewed
through a finite aperture pupil due to temperature-related
fluctuations in the refractive index of the atmosphere.  As the
spatial refractive-index fluctuations in the turbulent atmosphere move
between the telescope and the star at the wind speed, the wavefront is
perturbed, and the amount of starlight received by the telescope will
vary.  According to Young's formulation \nocite{you67} (Young 1967) of
Reiger's theory of scintillation \nocite{rei63} (Reiger 1963),
\begin{equation}
s_{scint}=\frac{c_{rms}}{\overline{c}} = S_0 d^{-2/3} X^{3/2} e^{-h/h_0}
\Delta f^{1/2},
\end{equation}
where $s_{scint}$ is the fractional rms variation due to
scintillation\footnote{We will often express fractional rms variation
in units of mmag.  For a change in flux $\Delta F$ that is small
compared to the average flux $F$, we have $\Delta m = 2.5 \log
((F+\Delta F)/F) = (2.5/\ln 10)\ln (1+\Delta F/F) \approx 1.086
(\Delta F/F)$, where $\Delta m$ is the magnitude change corresponding
to $\Delta F$.  Therefore, a 0.1\% flux variation is approximately
equal to a magnitude change of 1 mmag.}, $c_{rms}$ is the rms
variation in the source counts, $\overline{c}$ is the mean value of
the number of source counts, $c$, from a series of measurements, $S_0$
is a constant that \nocite{you67} Young (1967) determined empirically
to be 0.09, $d$ is the mirror diameter in cm, X is the air mass, $h$
is the observatory altitude in meters, $h_0=8000$ m, and $\Delta f$ is
the frequency bandpass of time-series sampling rate in Hz\footnote{The
frequency bandpass scaling of $s_{scint}$ arises from the fact that
the power spectrum of scintillation is flat \nocite{rei63,you67}
(Reiger 1963; Young 1967), up to a high frequency cutoff.
Since the variance is proportional to the integral of the power, $s^2
\propto
\int_{f_1}^{f_2}P(f)df \propto f_2-f_1$, where $f_2 \propto 1/t_{int}$
is the Nyquist frequency and $f_1 \rightarrow 0$, the rms deviation
$s_{scint}$ is proportional to $t_{int}^{-1/2} \propto \Delta
f^{1/2}$.}  (i.e. $\Delta f= 1.0/t_{int}$, where $t_{int}$ is the
integration time for each exposure in seconds).  The expression for
the total formal photometric error for each star in a field is thus
\begin{equation} \label{eqn:sigmam2}
\sigma_m^2=\sigma^2_{CCD,m}+s^2_{scint}c_m^2,
\end{equation}
where $\sigma_{CCD,m}$ is $\sigma_{CCD}$ from equation (\ref{eqn:ccd}) 
for star m with $c_m$ source counts. 
 
\begin{figure}
\begin{center}
\epsfig{file=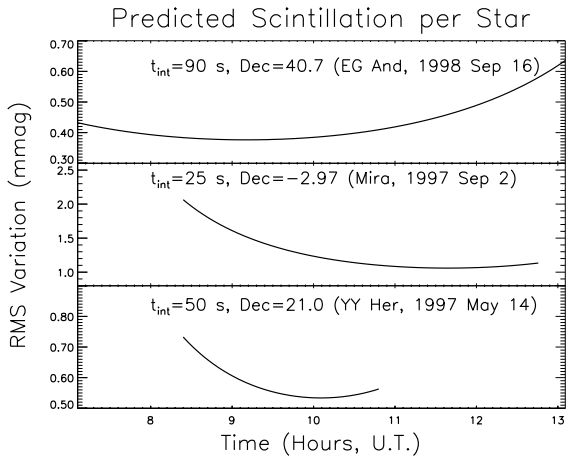,width=8cm}
\end{center}
\caption[Scintillation Contribution to RMS Variation.]{RMS
variation due to scintillation per star for three typical
observations.  The amplitude of the scintillation variability is a
function of integration time and airmass.  \label{fig:scint}}
\end{figure}

For the Nickel telescope at Lick Observatory, where $d=100$ cm and
$h=1280$ m, and scaling to typical values of air mass and integration
time, we find
\begin{eqnarray} \nonumber \label{eqn:scint}
s_{scint} & = & 0.84 \: {\rm mmag} \, \left(\frac{S_0}{0.09}\right) \left( \frac{d}{100\, {\rm
cm}}\right)^{-2/3} \left( \frac{X}{1.5} \right)^{3/2}  \\
 & & \hspace{1cm} \times \; e^{-h/1280\,{\rm m}} \left( \frac{t_{int}}{60\, {\rm s}} \right)^{-1/2}.
\end{eqnarray}
Assuming $S_0=0.09$ is applicable to our observations (this is a
conservative assumption, since whereas \nocite{you67} Young (1967)
measured this value for McDonald Observatory, \nocite{gb92} Gilliland
\& Brown (1992) inferred a value of 0.07 for Kitt Peak National
Observatory and speculated that even this value might be an upper
limit), we then find that even for the lowest airmass of 1.0, we still
have 0.5 mmag contribution to the rms variation from scintillation for
$t_{int}=60$ s.  The contribution climbs to 1.8 mmag for $X=2.5$ (and
$t_{int} =60\, {\rm s}$), which is comparable to the standard
deviation from Poisson statistics for our typical count rates.  Figure
\ref{fig:scint} shows the rms variation due to scintillation for
several typical observations.  Note that variability due to
scintillation is independent of the software extraction aperture size.

\subsection{Aperture Photometry and Construction of  Light Curves}

For the program star and as many comparison stars as possible (usually
2 to 6), the counts were extracted from circular apertures with radii
that typically ranged from 4.0 to 8.0 bins for CCD 2 and 5.0 to 9.5
bins for CCD 5 (3 to 6 arcsec in both cases).  The background was
estimated from an annulus surrounding the star of interest.  Very
faint comparison stars were not used, because for faint objects, the
increased number of data points that became unusable due to cosmic ray
contamination outweighed the small improvement in the S/N of the final
light curve.  Stars with blended images were also not used as
comparisons. We elected to use simple aperture photometry instead of
point-spread function (PSF) fitting for several reasons.  First, most
of our fields did not contain many stars, and the PSFs for these stars
were not well sampled due to our coarse binning.  Second, the PSFs
changed significantly as a function of time, so that a reliable PSF
would have been difficult to obtain.

For each image in a sequence of $N$ images used to produce a light
curve, the ratio of the program star counts to a weighted sum of
comparison star counts was found:
\begin{equation} \label{eqn:x}
x(i)=A\frac{c_p(i)}{\sum_{m=1}^{K}w_m c_m(i)},\;\;\;\;\;\;\;\; i=0, \ldots,N-1
\end{equation}
where $x(i)$ is the count ratio for the $i^{th}$ image, $c_p(i)$ is the
background subtracted source counts for the program star in the
$i^{th}$ image, $c_m(i)$ is the background subtracted source counts
for the $m^{th}$ comparison star in the $i^{th}$ image, $w_m$ is the
weight for the $m^{th}$ comparison star (same for every image, $i$), $K$
is the number of comparison stars used, and A is a normalization factor,
\begin{equation}
A^{-1}=\frac{1}{N}\sum_{j=0}^{N-1}\frac{c_p(j)}{\sum_{n=1}^K w_nc_n(j)}.
\end{equation}
The weights are given by
\begin{equation}
w_m = \overline{c_m} \, / \, \overline{\sigma_m^2} =
\frac{\sum_{i=0}^{N-1} c_m(i)}{\sum_{i=0}^{N-1}
\sigma_m^2(i) },
\end{equation}
where $\sigma_m$ is the formal uncertainty on $c_m$, the source counts
for star $m$, from equation (\ref{eqn:sigmam2}), and $\overline{c_m}$
and $\overline{\sigma_m^2}$ are the averages of $c_m$ and $\sigma_m^2$
over all $N$ images.  This form of the weights can be derived by
minimizing the uncertainty in the count ratio $x(i)$ (given in
equation (\ref{eqn:sigx})).  Note that the weights go to 1 in the
Poisson limit of $\sigma^2_m = c_m$.  This weighting scheme is the
same as that used by \nocite{gb88} Gilliland
\& Brown (1988), except that no position or color information is
included.  The series of $x$ values has been normalized to unity, so a
variation of 0.001 in $x$ corresponds closely to 1 mmag.  We refer to
the series of $x$ values as the light curve,
and individual $x$ values as data points.  The formal uncertainty for
each data point can be expressed as a function of the uncertainties
for the individual program and comparison stars:
\begin{equation} \label{eqn:sigx}
\sigma^2_x(i) \approx \left( \frac{\sigma_x(i)}{x(i)} \right)^2 = \left
( \frac{\sigma_p(i)}{c_p(i)} \right)^2 + \frac{\sum_{m=1}^{K} (w_m
\sigma_m(i))^2}{ \left[\sum_{n=1}^{K} w_n c_n(i) \right]^2 }.
\end{equation}
The variance is approximately equal to the fractional variance in the
expression above because, for the data presented here, the $x(i)$ are
all very close to unity.

\begin{figure}
\begin{center}
\epsfig{file=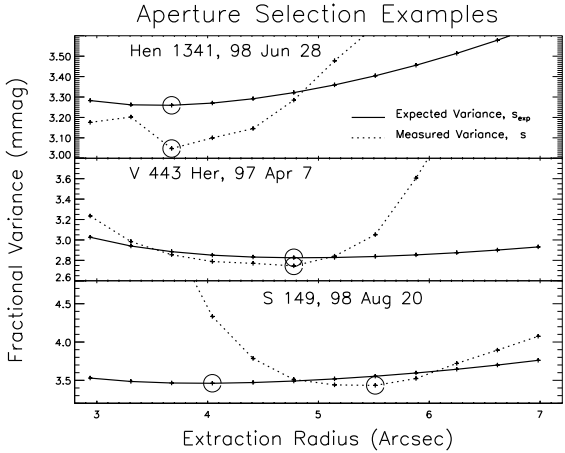,width=8cm}
\caption[Aperture Selection]{Three examples of the expected and measured
rms variation as a function of the photometry aperture.  In the first
two cases, the aperture which produced the lowest expected variance
also gave the lowest measured variance.  The aperture used for most
observations was that which produced the lowest expected variance,
even if that aperture did not produce the lowest measured variance.
The third panel shows an example of a case where systematic errors
were determined to be present at small apertures, so a larger aperture
was used.  In each of these examples, 4 comparison stars were used in
the construction of the light curves. \label{fig:apersel}}
\end{center}
\end{figure}

In constructing the final light curve, the aperture selected was
generally the one that maximized the S/N, or in other words, minimized
the expected variance $s_{exp}^2$, where $s_{exp}^2$ is a weighted
average of the $\sigma_x^2$ over all the points in a light curve,
\begin{equation} \label{eqn:sexp}
\sexp^2 = \frac{\sum_{i=0}^{N-1} \left( \frac{1}{\sigma_x^2(i)}\sigma_x^2(i)
\right)}{\sum_{i=0}^{N-1} \frac{1}{\sigma_x^2(i)}} =
\frac{N}{\sum_{i=0}^{N-1} \frac{1}{\sigma_x^2(i)}}.
\end{equation}
On the few occasions when the aperture which produced the best formal
S/N was very small (radius less than about 4 bins, or 3$\arcsec$ for
CCD 2), a slightly larger aperture was actually used to avoid problems
with pixelization noise or other systematic uncertainties if these
systematic errors were seen to be present in the comparison stars at
small extraction radii.  Pixelization noise was only a problem for
small apertures (see section
\S\ref{sec:error}).  A single aperture was used in extracting counts for
the program and all comparison stars throughout an entire light curve.
The use of different apertures for different parts of the light curve
would improve the formal S/N, as the background generally changed
throughout an observation.  However, we chose not to do this, as it
would introduce a new source of error because the PSFs are not quite
the same for the different stars in the field.  For observations with
a low background count rate, using a larger aperture increased the
source counts extracted without significantly increasing counts from the
background, and therefore improved the S/N.  For observations with a
high background rate, on the other hand, smaller apertures reduced the
uncertainty introduced by the background light and therefore produced
higher S/N light curves.  A few typical plots of $s_{exp}$ as a
function of aperture, as well as the measured rms variation, are shown
in Figure \ref{fig:apersel}.

\begin{figure}
\begin{center}
\epsfig{file=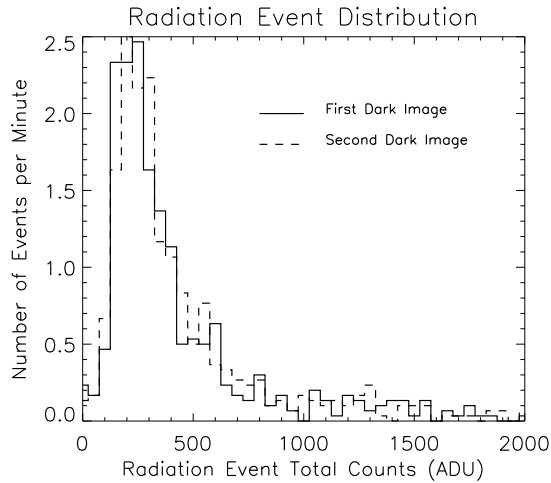,width=8cm}
\caption[Cosmic Ray Energy Distribution]{Energy distribution of cosmic rays, from two
30-minute dark images taken with CCD 2. \label{fig:crs}}
\end{center}
\end{figure}

Data points contaminated by radiation events (``cosmic rays'') were
omitted, as were points with extremely high background, or taken
during exceptionally poor weather conditions.  Data points
contaminated by cosmic rays were identified in two ways.  First, the
residuals from Gaussian fitting were compared for the various stars in
a field.  This method worked well for cosmic rays in the wings of a
PSF, but less well for cosmic rays that hit near the center of the
PSF. Next, light curves were formed for all possible pairs of
comparison stars in a field.  If any point in these light curves was
suspiciously high or low (by greater than at least 5 $\sigma$,
although usually much more), that point was not used in the creation
of the final ensemble average.  Occasionally, individual extreme high
points were due to a cosmic ray hit near the program star itself, and
these points were also removed.  We were much more conservative,
however, about removing points in which the program star was the one
thought to be contaminated.  In general, only about 5 to 10 points per
light curve were rejected because of cosmic ray contamination.  This
number is consistent with the expected rate of cosmic ray hits, given
the measured distribution of cosmic rays from dark images.  We
measured the rate of cosmic ray hits to be $1.25\times 10^{-6} \,{\rm
s}^{-1}{\rm bin}^{-1}$ for CCD 2.  The number of contaminated data
points per observation, $n_{crs}$, is then expected to be
\begin{equation}
n_{crs}\approx 5 \left(\frac{K+1}{5}\right) \left(\frac{N}{120}\right)
\left(\frac{6}{b} \right)^2 \left(\frac{t_{int}}{60 \,{\rm s}} \right),
\end{equation}
where $K$ is the number of comparison stars, $N$ is the number of data
points, $b$ is the extraction radius in bins, and $t_{int}$ is the
integration time (note that radiation events can also occur during
read-out).  The distribution of total counts in a cosmic-ray event,
measured from two 30-minute dark images, is plotted in Figure
\ref{fig:crs}.  The most typical event size was approximately 200 ADU
$\approx$ 1400 $e^-$.  Since a typical number of source counts was on
the order of $10^5$, an unidentified cosmic ray within the stellar
aperture could therefore artificially increase the source flux by
roughly 1\%.

Finally, the resulting light curve was fit with a 2nd or 3rd-order
polynomial, depending upon the length of the observation, to remove
any slow variations due to changing air mass.  These variations could
exist, despite using differential photometry, when the program star
and comparison stars had significantly different colors.

\section{Timing Analysis}\label{sec:timing}

The survey light curves were analyzed in several ways.  First, the
measured rms variation was compared to that expected from the CCD
equation and scintillation.  Next, power spectra were computed to
search for periodic signals, and Monte Carlo simulations performed to
determine our detection limits for sinusoidal variations in each light
curve.

\subsection{Aperiodic Variability}

\begin{figure}
\begin{center}
\epsfig{file=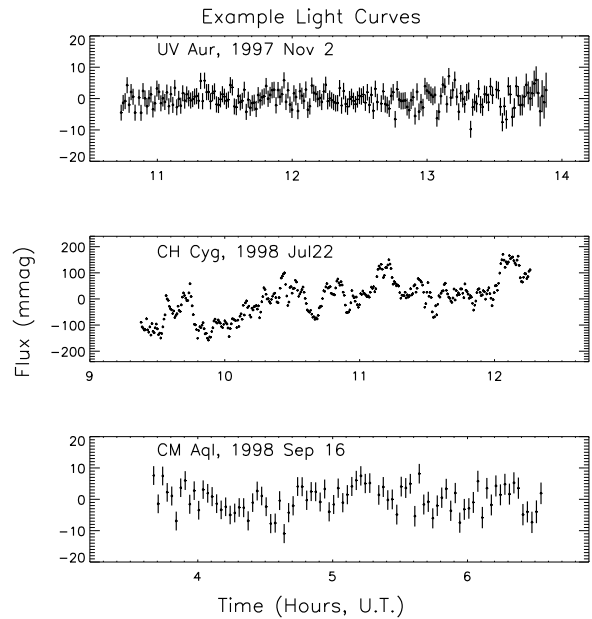,width=8cm}
\end{center}
\caption[Example Light Curves]{Example light curves showing, from top
to botttom, a
non-detection, a clear detection, and a marginal detection of rapid
variability.  \label{fig:lcexs}}
\end{figure}

The crux of our search for aperiodic variability, or flickering, was
comparing the measured rms variation for each light curve (denoted
$s$) to the rms variation expected from the known errors, denoted
$\sexp$, defined in equation (\ref{eqn:sexp}).  The measured rms
variation is obtained from the data using the general expression for
the variance when the errors on each data point are
different (\nocite{bev69}Bevington 1969, p. 185),
\begin{equation}
s^2 = \frac{N}{N-1} \frac{\sum_{i=0}^{N-1}
\left(\frac{x_i-\overline{x}}{\sigma_i} \right)^2}{\sum_{j=0}^{N-1}
\frac{1}{\sigma_j^2}}.
\end{equation}
Note that comparing the two quantities $s$ and $\sexp$ is actually the
same as calculating $\chi^2$ for the hypothesis of a constant flux
(\nocite{bev69}Bevington 1969, p. 188).  If $s$ and $\sexp$ are
comparable (i.e., $s/\sexp \approx 1$), the fit of the data to a
constant flux model is good, but if $s/\sexp$ is significantly greater
than 1, the fit is bad and the source is possibly variable.  Plotted
in Figure \ref{fig:lcexs} are examples of three different types of
light curves.  In the top panel (UV Aur), $s/\sexp \approx 1$, and we
find a variability upper limit of $s = 2.7$ mmag.  Most light curves
in our survey fall into this category, and a list of the variability
upper limits is given in Table \ref{tab:upperlims}.  In the middle
panel, we present the other extreme --- a light curve of the
well-known flickerer CH Cyg (e.g., \nocite{mski90a} Miko{\l}ajewski et
al 1990a, \nocite{hoa93} Hoard 1993,
\nocite{rhbmor97} Rodgers et al. 1997, \nocite{eim98} Ezuka et
al. 1998).  For this light curve we find $s/\sexp = 19.4$, an
unequivocal detection of variability.  We did not find any new objects
with this level of variability.  An analysis of the variability
properties of the five previously-known, large-amplitude flickerers is
given in
\nocite{sbh01} Sokoloski et al. (2001b).  The bottom panel of Figure
\ref{fig:lcexs} shows a marginal detection of flickering in the light
curve of survey object CM Aql.  In this case, $s/\sexp = 1.4$, and we
are at or near our limits of detectability.  Because of its small
amplitude, this variability in this light curve could possibly be due
to some unidentified systematic error.  Measured and expected
variances for the detections and marginal detections of variability
are listed in Table \ref{tab:candvars}.

\subsection{Periodic Variability} \label{sec:pdvar}

The main tool we used to search for periodic variability was Fourier
analysis of the light curves.  In the discussion that follows, we
consider a light curve that is simply a series of source count values
(i.e., not ratios of source and comparison star count values), but the
results apply for differential light curves as well.  Our observations
consisted of data points that were at least approximately evenly
spaced, so we were able to use a simple discrete FFT routine for this
analysis.

Consider a power spectrum constructed using the \nocite{lea83} Leahy
et al. (1983) normalization:
\begin{equation} \label{eqn:leahy}
P_j \equiv \frac{2}{C_{tot}} |a_j|^2,\;\;\;\;\;\;\;\;j=0, \ldots, \frac{N}{2},
\end{equation}
where $C_{tot} = \sum_{k=0}^{N-1} c(k)$ is the total number of source
counts in a light curve ($c(k)$ is the number of source counts in
exposure $k$) and the $a_j$ are the Fourier amplitudes for the set of
frequencies $\nu_j = j / (N \Delta t)$ ,
\begin{equation}
a_j = \sum_{k=0}^{N-1} c(k) e^{2\pi ijk/N} \;\;\;\;\;\;\;\;j=-\frac{N}{2},
\ldots, \frac{N}{2}-1
\end{equation}
($\Delta t$ is the time between data points).  Note that $a_0 =
C_{tot}$.  With this normalization, the average value of the power
from from a time series of purely Poison noise is 2 (see Appendix A).
If a light curve consists of a constant plus pure Poisson noise, then
the underlying noise in the power spectrum normalized as defined above
is distributed like $\chi^2$ with 2 degrees of freedom
(\nocite{vdk89}van der Klis 1989; see \nocite{gro75} Groth 1975 for
derivation).  One can therefore integrate the $\chi^2$ distribution to
find the probability that a noise power will exceed the power level
$P$ in a given frequency bin:
\begin{equation} \label{eqn:prob}
\Pr (P_{j,noise} > P) = \frac{1}{2} \int_P^\infty e^{-y/2} dy = e^{-P/2}.
\end{equation}
Many other types of noise also produce a $\chi^2$ distribution of the
noise powers \nocite{vdk89} (van der Klis 1989).  In the general case of single or
multiple noise sources which are normally distributed, 
the distribution function of noise powers can be written as
\begin{equation}
f(y)=\frac{1}{\overline{P}}e^{-y/\overline{P}},
\end{equation}
so that a more general expression for the probability that a noise
power will exceed the power level $P$ in a given frequency bin is
\begin{equation} \label{eqn:onebinprob}
\Pr (P_{j,noise} > P) = \frac{1}{\overline{P}} \int_P^\infty
e^{-y/\overline{P}} dy = e^{-P/\overline{P}}
\end{equation}
(\nocite{gro75,sca82}Groth 1975; Scargle 1982), where $\overline{P}$ is the average noise power
near the frequency $\nu_j$.  

When examining an entire power spectrum, however, we are no longer
asking about the chance of a noise power exceeding a given power $P$
in a particular frequency bin, but instead the chance of a noise power
exceeding $P$ in any bin.  Since the probability that the power in a
given frequency bin will not reach the level P is
$1-e^{-P/\overline{P}}$, the probability that none of the points in
the power spectrum will reach power P is
$(1-e^{-P/\overline{P}})^{n_{freq}}$, where $n_{freq}$ is the number
of frequency bins.  The analogous expression to equation
(\ref{eqn:onebinprob}) for a multiple frequency bin search is
therefore
\begin{equation} \label{eqn:mfreqsearch}
\Pr (P_{noise} > P) = 1 - \left(1 -e^{-P/\overline{P}} \right)^{n_{freq}},
\end{equation}
where $\Pr (P_{noise} > P)$ is the probability that at least one of the
noise powers will exceed $P$.  Note that in the limit where
$e^{-P/\overline{P}}$ is small, this probability for obtaining a noise
power greater than $P$ approaches the more intuitive $n_{freq} \times
e^{-P/\overline{P}}$.  

If a power spectrum computed from one of our survey light curves
contained a peak that had less than a 5\% chance of having been
produced by statistical noise, then that observation was flagged as a
potential detection of periodic variability.  In other words, if $\Pr
(P_{noise} > P) < 0.05$, then there was a 95\% chance that a peak in a
power spectrum was due to intrinsic source variability.  By choosing a
confidence level of 95\% for the detection threshold, our set of light
curves could have produced several spurious detections.  This fairly
low detection threshold was selected so that real signals were
unlikely to be missed, and because follow-up observations could be
performed fairly easily (e.g. as were done for Z And; see Sokoloski \&
Bildsten 1999).

Whereas one can easily evaluate the likelihood that a detected
power-spectrum peak is due to intrinsic source variability using the
expressions above, the absence of a statistically significant peak is
slightly more difficult to interpret.  To see what sort of upper
limits we can place on the amplitute of a periodic signal, if no
significant power-spectrum peak is present, we calculate our
sensitivity below, based on the treatment of this subject in
\nocite{vdk89} van der Klis (1989).  

The smallest-amplitude signal to which we are sensitive corresponds to
a power $P_{sensitive}=P_{det}-P_{exceed}$, where $P_{det}$ is the
minimum power for a secure detection of true source variability, and
$P_{exceed}$ is a low value that is very likely to be exceeded by a
noise power.  Setting the probability in equation
(\ref{eqn:mfreqsearch}) to a small number $\epsilon$, we get
$P_{det}=-\overline{P} \,\ln [1-(1-\epsilon)^{1/n_{freq}}]$.  Setting
the probability in equation (\ref{eqn:onebinprob}) to a large value
$1-\delta$, gives $P_{exceed} \equiv -\overline{P}
\,\ln (1-\delta)$.  Putting these two expressions together, we find
\begin{equation}
P_{sensitive} = \overline{P} \,\ln \left [
\frac{1-\delta}{1-(1-\epsilon)^{1/n_{freq}}} \right],
\end{equation}
with $1-\delta$ and $1-\epsilon$ both taken to be large (i.e., close
to 1).  It can be shown that the amplitude, $A$, of a sinusoidal
signal is related to the square of the Fourier amplitude for the bin
containing the signal frequency by the expression
\begin{equation}
|a_j|^2 = \alpha \frac{1}{4} A^2 N^2,
\end{equation}
where $\alpha$ has an average value of 0.77 and depends upon the
location of the signal frequency in the frequency bin
(\nocite{vdk89}van der Klis 1989).  Substituting $|a_j|^2 =
(P_{sensitive}C_{tot})/2$ (from equation \ref{eqn:leahy}, for the
Leahy normalization) and solving for $A$, we find
\begin{equation} \label{eqn:sens}
A_{sensitive} \approx 1.14 \frac{\sqrt{C_{tot} }}{N} \ln
\left[\frac{1-\delta}{ 1-(1-\epsilon)^{1/n_{freq}}} \right]. 
\end{equation}
Plugging in some typical numbers and taking $1-\delta=1-\epsilon=0.95$,
we find that in theory we are sensitive to sinusoidal signals with a
fractional amplitude
\begin{equation}
\frac{A_{sensitive}}{\overline{c}} \approx 0.001 \left
( \frac{C_{tot}}{5\times 10^7} \right)^{-1/2},
\end{equation}
for $N=200$ and $n_{freq} = 100$ (recall that $\overline{c}$ is the
mean value of the number of source counts, $c$, from a series of
measurements).  This fractional amplitude is roughly equal to 1 mmag.

Since the power of a sinusoidal signal with a given amplitude depends
upon the location of the signal frequency within a frequency bin, the
signal amplitude to which we are sensitive is only approximately given
by the expression above.  Thus, the best way to determine a signal
amplitude upper limit when there are no significant peaks in the power
spectrum is through simulations.  For each light curve, we therefore
added sinusoidal variations with a range of amplitudes, frequencies,
and phases to the data, and computed the power spectra.  The trial
frequencies ranged from the Nyquist value to $3/T_{tot}$ for each
light curve, where $T_{tot}$ is the total length of a light curve.
Using $10^4$ realizations for each trial amplitude (100 trial
frequencies, and 100 phases for each frequency), we took the signal
amplitude upper limit to be that which produced a 95\% confidence
level detection 95\% of the time.  The periodic variability upper
limits for each object for which no significant peaks were present in
the power spectrum are listed in Table \ref{tab:upperlims}.

\subsection{Sources of Error} \label{sec:error}

To determine the amount of systematic error in our differential
photometry, we examined the comparison stars in multiple fields to
determine whether there was a systematic excess variance beyond that
expected from the CCD equation and scintillation.  For star $m$ in
image $i$ (out of $N$ images in an observation), the expected
fractional variance is given by
\begin{equation}
s_{exp,m}^2(i) = \left( \frac{\sigma_{CCD,m}(i)}{c_m(i)} \right)^2 + s_{scint}^2,
\end{equation}
where $\sigma_{CCD,m}^2(i)$ is the variance from equation (\ref{eqn:ccd})
for star $m$ in image $i$, $c_m(i)$ is the number of stellar
counts for star $m$ in image $i$, and $s_{scint}$ is the scintillation 
given by equation (\ref{eqn:scint}).  If the ratio of the fluxes of two
stars $l$ and $m$ is formed to remove the effects of clouds and seeing
variations, the expected fractional variance of this ratio is
\begin{equation} \label{eqn:ratvar}
s_{exp,lm}^2(i) = s_{exp,l}^2(i) + s_{exp,m}^2(i).
\end{equation}
The average expected fractional variance for this flux ratio
for the entire observation is
\begin{equation}
s_{exp,lm}^2=\frac{\sum_i \left( \frac{1}{s_{exp,lm}^2(i)} \
s_{exp,lm}^2(i) \right) }{\sum_j 1/s_{exp,lm}^2(j)} = 
\frac{N}{\sum_j 1/s_{exp,lm}^2(j)}. 
\end{equation}
For every possible pair-wise combination of comparison stars in the
test fields, the flux ratio was formed, the expected fractional
variance from the CCD equation and scintillation was calculated, and
the actual fractional variance of the flux ratio was measured.  By
solving a simple system of equations that had the same form as
equation (\ref{eqn:ratvar}), we could then obtain the variance due to
systematic errors for each star individually.

In most cases, little systematic error was found. Furthermore, the
examination of measured versus expected rms variation for
constant-brightness field stars allowed us to come up with a criterion
for identifying low-level, flickering-type source variability.  The
constant comparison stars generally had variances that were within
40\% of the expected values (both measured and expected variances were
typically on the order of a few mmag).  Therefore, any symbiotic in
the survey with $s/s_{exp}
\ge 1.4$ was considered to have a potential detection of variability.

Looking at the comparison-star fluxes with respect to each other in
each field also allowed us to identify any variable comparison stars
and exclude them from the analysis. Data from a new pulsating A-type
star with a 30-minute period that was discovered in this way, as well
as several other serendipitous discoveries, are presented in
\nocite{sbcf01} (Sokoloski et al. 2001a).

\begin{figure}
\begin{center}
\epsfig{file=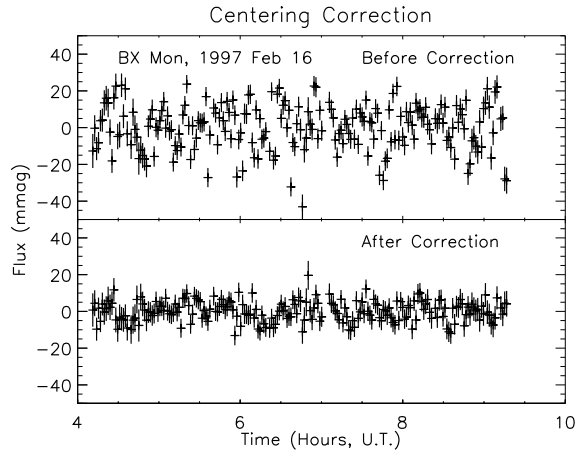,width=8cm}
\end{center}
\caption[Centering Correction]{A particularly extreme example of the
improvement in our light curves once a centering correction was applied.
These light curves were constructed from the 8 comparison stars in the
field of BX Mon on 1997 Feb 16, using an aperture radius of 2.9$\arcsec$.
\label{fig:censcat}}
\end{figure}

As mentioned above, we found that in most cases, systematic error was
negligible.  However, there were a few instances in which problems
arose.  The first problem we found was a result of the poor
centroiding performed in the IRAF-based IDL routine ``cntrd''.  The
centroiding problems are evident from the large scatter in the
difference between stellar positions output from ``cntrd'' for a
series of images of the same field.  Improved centers were obtained by
assuming the PSFs had a roughly bivariate Gaussian shape, and fitting
the PSFs with this function with the $x$ and $y$ centers as free
parameters.  Figure
\ref{fig:censcat} shows a differential light curve for comparison
stars in the field of BX Mon on 1997 Feb 26, first before, and after
the centering improvement.

\begin{figure}
\begin{center}
\epsfig{file=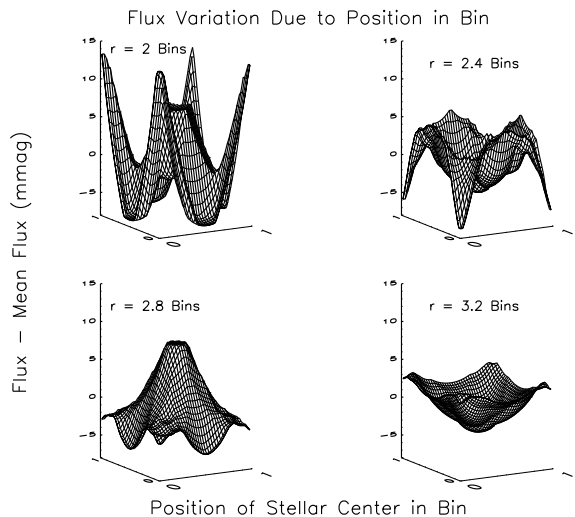,width=8cm}
\end{center}
\caption[Flux Variation Due to Position in Bin]{Four examples of the
extracted flux as a function of position of the stellar PSF center in
a bin.  Plotted is the difference in flux from the mean value in units
of mmag, for a Gaussian point-spread function with $\sigma_{Gauss}$ =
1.4 bins.  Note that the variation is smaller for larger extraction
radius, $r$.
\label{fig:pix1}}
\end{figure}

\begin{figure}
\begin{center}
\epsfig{file=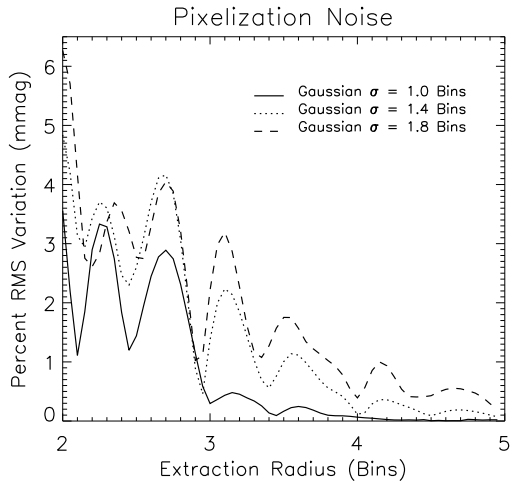,width=8cm}
\end{center}
\caption[RMS Variation Due to Pixelization Noise]{Simulated
``pixelization noise'' for a range of PSF widths and extraction radii.
For CCD 2, 1 bin is equal to 0.74$\arcsec$, and for CCD 5, 1 bin is
equal to 0.59$\arcsec$.  A point spread function with $\sigma_{Gauss}
= 1.4$ bins on CCD 2 therefore corresponds to ``seeing'' of about 2
$\arcsec$.
\label{fig:pix2}}
\end{figure}

A second source of error, ``pixelization noise'', is only relevant for
small extraction apertures.  This error arises because of imperfect
estimation of the fraction of counts that should be included from a
bin that falls only partially within the extraction region.  For large
apertures, there are relatively few edge bins compared to bins that
are fully within the aperture, and the effect of pixelization is
small.  To estimate the magnitude of pixelization noise for various
aperture sizes, and for stellar PSF profiles of various widths, we
created very finely sampled Gaussian stellar profiles, and binned them
as they would be on our CCDs, with the true center of the profile
placed at $2.5\times 10^5$ different positions in a bin.  We then
performed circular-aperture photometry with our software for each
different PSF position.  In Figure \ref{fig:pix1}, several contour
plots show the change in extracted flux as a function of position of
the PSF center within a bin.  In Figure \ref{fig:pix2} is plotted the
fractional variation in the extracted fluxes as a function of
extraction aperture for various Gaussian profile widths.

Furthermore, there appears to be a slight positional dependence of the
PSF (at least for CCD 2), which can be seen by comparing the $x$ and
$y$ width of the Gaussian fits for the different stars in a field.
For observations with stable seeing, this PSF difference is not a
problem.  If the seeing changed significantly during an observation,
however, then one of the assumptions of our differential photometry
was violated, and the light curve could exhibit variability that was
either directly or inversely correlated with seeing.  It is also
possible that excess variance in the final ensemble average light
curve could be due to some cosmic rays within the program star
extraction region.  We were more cautious about removing data with
suspected cosmic ray hits on the program star than we were with data
in which the comparison stars were affected, in order to avoid
removing real variability.  However, the numerous light curves for
which we do not find excess variability in the program star indicate
that this is probably not a significant issue.

In a few observations, a correlation between the differential light
curve and a measure of the seeing, the position of the stars on the
chip, features seen in the raw light curves of all stars in the field,
the background, or some other house-keeping information indicated the
possible presence of some systematic error.  Likely candidates for the
source of the error include differences in the PSFs, blended images,
differential extinction, or positional dependences of gain or bias
features (see \nocite{gb88}Gilliland \& Brown 1988, who find that
color and position corrections matter).  Any data that were clearly
affected by systematic error (usually during extremely poor
weather) were not used, and some questionable cases are discussed
later.  For most observations, however, all of the above effects were
small compared to statistical noise sources such as Poisson noise due
to counting statistics and read noise.

\section{Results}\label{sec:sresults}

\subsection{Upper Limits}

\begin{figure*}
\epsfig{file=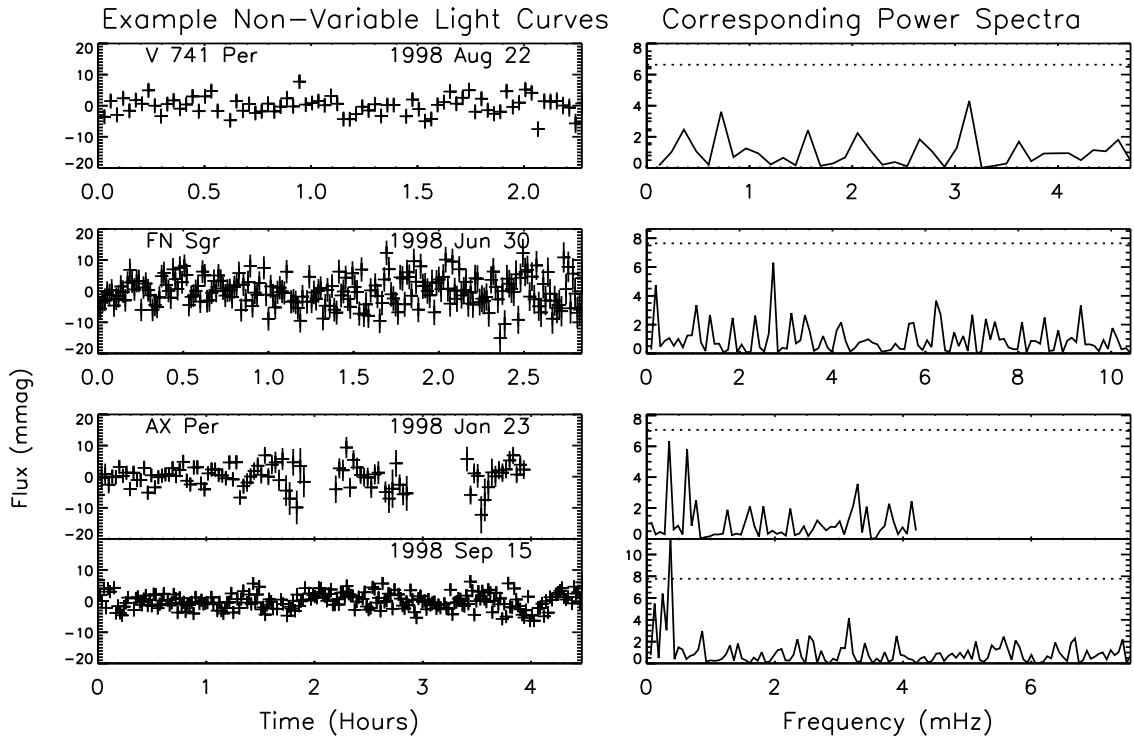,width=16cm} 
\vspace{-1.5cm}
\caption[Example Light Curves of Survey Objects with No Detected Rapid
Variability.]{Example light curves and power spectra for several
non-variable survey objects. All the non-variable objects are listed,
with their variability upper limits, in Table \ref{tab:upperlims}. The
flux is in units of mmag, and the observations were done with a $B$
filter.  The power is normalized by the mean value for an observation,
and the dotted line in the power-spectrum plots indicates the power
needed for a 95\% confidence level signal detection.  For the 1998 Sep
15 observation of AX Per, a peak in the power spectrum rises above the
formal detection level, but we do not believe this peak is due to
intrinsic source variability.  This type of observation is discussed
in the text. \label{fig:somelcs}}
\end{figure*}

For 25 of the 35 objects in our survey, no variability,
either periodic or aperiodic, was detected.  The fractional rms
variations ($s$) for these objects are listed in Table
\ref{tab:upperlims}.  Some representative light curves and corresponding power spectra are
shown in Figure \ref{fig:somelcs}.  The value of $s$ for these 25
objects ranged from 2.4 to 10.4 mmag, giving $s/s_{exp} \le 1.4$
(where $s_{exp}$ is the rms fractional variation expected from Poisson
statistics and scintillation, defined in equation \ref{eqn:sexp}) for
all but one observation, which we discuss below.  The upper limits on
the amplitude of any periodic oscillations present in these data fall
in approximately the same range of a few to 10 mmag, or less than a
1\% flux variation (see Table
\ref{tab:upperlims}).  As discussed in \S
\ref{sec:pdvar}, the oscillation-amplitude upper limits correspond to
the amplitude of a sinusoidal signal which, when injected into the
data at a range of periods and phases, produced a formal 95\%
confidence detection 95\% of the time.

Two of the observations presented in Table \ref{tab:upperlims} as
upper limits actually contained some low-level variability, which we
do not believe to be due to intrinsic source variations.  The power
spectrum from the 1998 Sep 15 observation of AX Per has a peak above
the formal signal-detection level.  However, significant low-frequency
power was also seen in some of the comparison stars as well as AX Per,
and so systematic effects were probably present.  In the 1997 Jun 7
observation of AG Dra, which was done with only a single comparison
star, the flux appears related to the position of the star on the
chip, and the variation is therefore also believed to be due to a
systematic effect.  In both cases, a second observation showed no
variability.

The main result for the 25 objects listed in Table
\ref{tab:upperlims}, and therefore one of the main results from our
35-object survey, is that although most SS contain accreting WDs, as
in cataclysmic variables, they generally do not show CV-like optical
flickering.  For over 70\% of the symbiotics in this survey, the
amplitude of any rapid optical variability is constrained to just a
fraction of a percent.

\setcounter{table}{3}
\begin{table*}
\centering
\begin{minipage}{130mm}
\caption[Variability Upper Limits]{Variability Upper Limits\label{tab:upperlims}}
\begin{tabular}{lcccccc}
\\
\hline
\hline
\\
Name & Date & $s_{exp}$ & $s$ & & Max Sine & Period Range \\ 
 & (m/d/yr)  & (mmag) & (mmag) & & Amplitude (mmag) & (min) \\
\\
\hline 
\\
AX Per & 1/23/98 & 2.8 & 3.2 & & 4.1 & 3.9 - 78.6 \\*
       & 9/15/98\footnote{The timing errors are larger than usual for
observations done in 1998 Sep because the high precision WWVB clock
was broken.  For these observations, the time between integration starts is
uncertain by up to a few seconds.} & 1.9 & 2.6 & & 1.9 & 2.2 - 88.9 \\
V 741 Per & 8/22/98 & 2.4 & 2.8 & & 3.2 & 3.5 - 45.6 \\
S 32 & 3/13/97\footnote{On 1997 Mar 13, incorrect time stamps were recorded in
the image headers.  The timing for this date is therefore only approximate.} & 7.0 & 8.2 & & 13 & 3.1 - 27.0 \\*
     & 9/14/98$^a$ & 2.7 & 3.2 & & 3.8 & 3.2 - 45.0 \\
UV Aur & 11/2/97 & 2.7 & 2.7 & & 1.9 & 1.8 - 62.7 \\
TX CVn  & 3/13/97$^b$ & 3.2 & 4.4 & & 4.7 & 2.9 - 63.5 \\*
RW Hya & 2/18/97 & 3.4 & 4.2 & & 3.6 & 1.8 - 41.2 \\*
BD-21.3873  & 5/14/97 & 3.1 & 3.5 & & 4.6 & 3.8 - 40.5 \\*
            & 5/31/98 & 4.7 & 6.1 & & 7.3 & 3.0 - 36.2 \\
AG Dra  & 6/7/97 & 3.9 & 6.3 & & 4.7 & 1.6 - 66.5 \\*
       & 7/23/98 & 5.4 & 5.8 & & 4.0 & 2.7 - 92.9 \\
HD 154791 & 8/4/97 & 9.7 & 10.4 &  & 4.6 & 1.1 - 96.9 \\*
Hen 1341 & 6/28/98 & 3.3 & 3.1 & & 3.1 & 3.1 - 46.6 \\
AS 289 & 5/31/98 & 6.2 & 7.7 & & 9.8 &3.7 - 51.9 \\*
YY Her & 5/14/97 & 5.4 & 5.3 & & 4.0 & 2.5 - 76.9 \\
AS 296 & 8/21/98 & 2.3 & 2.4 & & 2.5& 3.8 - 61.4 \\
S 149 & 5/15/97 & 3.5 & 3.8 & & 3.8 & 2.8 - 44.7 \\*
      & 8/20/98 & 3.6 & 3.4 & & 2.5 & 2.6 - 91.5\\
V 443 Her & 4/7/97 & 2.8 & 2.7 & & 2.5 & 2.6 - 55.2 \\
FN Sgr & 6/30/98 & 3.9 & 4.6 & & 3.3 & 1.6 - 56.5 \\*
V 919 Sgr & 8/22/98 & 3.0 & 3.0 & & 3.2 & 3.9 - 49.3 \\
CI Cyg & 6/8/97 & 3.7 & 4.0 & & 2.6 & 1.5 - 61.3 \\*
       & 9/15/98$^a$ & 3.9 & 5.4 & &5.8 & 4.2 - 87.5 \\
AS 360 & 6/29/98 & 3.4 & 4.8 & & 4.7 & 2.6 - 46.8 \\
PU Vul & 6/15/97 & 3.1 & 3.8 & & 5.1 & 2.0 - 27.4 \\
He2-467 & 8/22/98 & 2.5 & 2.4 & & 2.9 & 3.9 - 42.2 \\
S 180 & 7/23/98 & 5.5 & 5.5 & & 11 & 4.2 - 30.1 \\
S 190 & 8/21/98 & 3.7 & 4.5 & & 3.4 & 2.2 - 68.6 \\
AG Peg & 9/17/98$^a$ & 2.9 & 3.5 & & 7.3& 1.6 - 50.4\\
R Aqr & 9/14/98$^a$ & 4.6 & 5.4 & & 3.4 & 1.7 - 78.1 \\
\end{tabular}
\end{minipage}
\end{table*}

\subsection{New Candidate Flickerers} \label{sec:newcandflick}

In 4 systems, evidence for aperiodic rapid variability was found for the
first time.  We consider an observation to contain a potential detection
of rapid variability if $s/s_{exp} \simgt 1.4$, and the variations are
not correlated with any of our house-keeping data.  Since all of these
measurements are near our detection limits, however, they should be
considered rather uncertain, and further observations are needed to
confirm the presence of flickering in these objects.  The candidate
variable systems are discussed individually below.  Their expected and
measured fractional rms variations are listed in Table \ref{tab:candvars}.

\subsubsection{EG And} \label{sec:egand}

\begin{figure}
\begin{center}
\epsfig{file=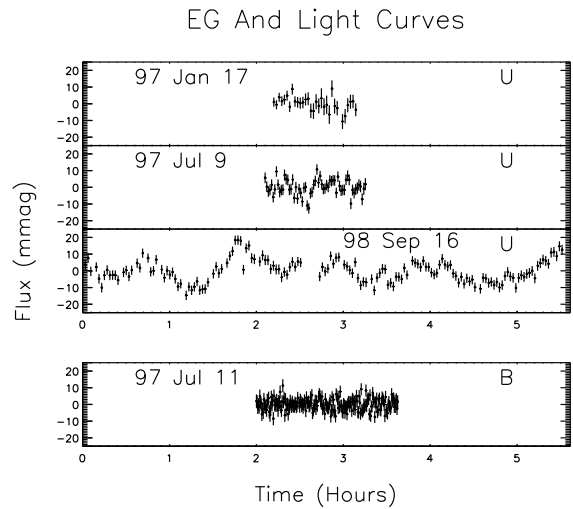,width=8cm}
\end{center}
\caption[EG And Light Curves.]{Light curves from the 4 observations of EG
And are shown.  Statistically significant variability is seen in the
second and third $U$-band light curves, whereas the $B$-band light
curve is consistent with an assumption of constant
flux. \label{fig:egandlcs}}
\end{figure}

EG And was observed on 4 occasions, 3 times with a $U$ filter, and
once with a $B$ filter.  In 2 of the $U$-band observations, on 1997
Jul 9 and 1998 Sep 16, there is evidence for low-level flickering.
The measured variance was 50\% above the expected level on 1997 Jul 9,
and more than twice the expected level on 1998 Sep 16.  The
observation that produced the most variable light curve, on 1998 Sep
16, however, was performed using the smaller-field CCD 5, which only
allowed for one reasonably bright comparison star.  Moreover, this one
bright comparison star was near a feature on the chip which introduced
a slow rise in the comparison star flux as the star drifted toward the
feature.  There does not appear to be a correlation between the
position of the comparison star and the rapid flux variations in the
light curve, but the presence of this chip feature does introduce
another reason for caution when interpreting these results.  Plots of
the four EG And light curves are shown in Figure \ref{fig:egandlcs}.
The power spectrum from 1998 Sep 16 contained significant
low-frequency power spread over several frequencies.  No
statistically-significant peaks were present in the power spectra from
the other three observations.

\begin{figure}
\begin{center}
\epsfig{file=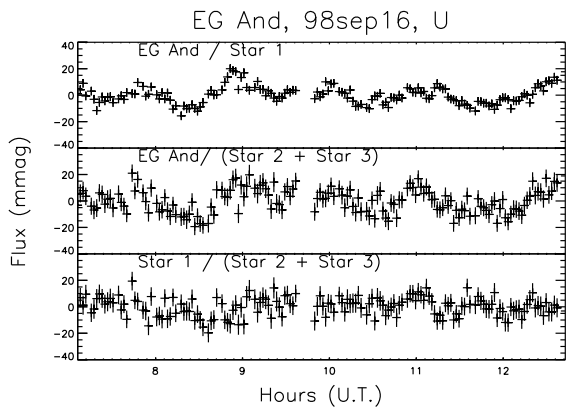,width=8cm}
\end{center}
\caption[Flux Ratios for EG And.]{Flux ratios for EG And and other
stars in the same field.  For a secure detection of variability in EG
And, the light curves in the first 2 panels should show the same
trends, while the light curve in the third panel should not. Based on
these light curves, we classify this observation as a marginal
detection of variability in EG And.  \label{fig:egand916rats}}
\end{figure}

To illustrate the difficulty with interpretation of these
observations, Figure \ref{fig:egand916rats} shows, for 1998 Sep 16,
the flux ratio of EG And flux versus the main comparison star, the
flux ratio of EG And flux versus the sum of 2 very faint star in the
field, and the flux ratio of the main comparison star versus the sum
of the 2 faint stars.  For a high-confidence-level detection, one
would like to see significant and correlated variability in both the
first and second ratios, but not the third. The measured variance is
in fact more than twice the expected variance ($s/\sexp=2.4$) for the
first ratio, the measured variance is approximately $40\%$ higher than
that expected from a constant source for the second
ratio($s/\sexp=1.4$), and the 2 quantities are nearly the same for the
third ratio ($s/\sexp=1.1$).  For this object, it is also interesting
to note that the very flat $B$ light curve of 1997 Jul 11 is only 2
nights away from a $U$ observation that shows hints of variability.

\begin{table*}
\caption[Candidate Variable Systems ]{Candidate Variable Systems \label{tab:candvars}}
\begin{tabular}{lcccccccc}
\hline
\hline
\\
 & & & \multicolumn{3}{c}{Fractional Variation} & &
\multicolumn{2}{c}{Power Spectrum} \\
\cline{4-6} \cline{8-9} \\
Name & Date & Filter & $s_{exp}$ & $s$ & $s/s_{exp}$ & & Max. Sine & Period Range \\ 
 & (m/d/yr) & & (mmag) & (mmag) & & & Amplitude (mmag) & (min) \\
\\
\hline 
\\
EG And & 1/17/97 & $U$& 3.7 & 3.8 & 1.0 & & 6.9 & 3.6 - 19.3 \\*
       & 7/9/97 & $U$&3.2 & 4.7 & 1.5 & & 5.8 & 2.2 - 23.2 \\* 
       & 7/11/97 & $B$ & 3.7 & 3.5 & 0.9 & & 2.3 & 0.9 - 32.3 \\* 
       & 9/16/98 & $U$& 2.8 & 6.7 & 2.4 & &
\multicolumn{2}{c}{Low-frequency power present} \\ 
BX Mon & 1/18/97$^b$ & $B$ & 10.2 & 9.1 & 0.9 &  & - & - \\* 
       & 2/16/97 & $B$ & 3.7 & 9.9 & 2.7 & &
\multicolumn{2}{c}{Low-frequency power present}\\*
       & 3/12/97 & $U$& 6.8 & 13.1 &1.9 &
&\multicolumn{2}{c}{Low-frequency power present} \\*
       & 4/7/97 & $B$ & 4.5 & 6.0 & 1.3 & & 10.2 & 2.8 - 19.7 \\*
       & 11/1/97 & $B$ & 3.9 & 4.2 & 1.1 & & 4.3 & 3.4 - 51.1\\*
CM Aql & 9/16/98$^a$ & $B$ & 3.1 & 4.2 & 1.4 & &4.3 & 4.0 - 57.6 \\
BF Cyg & 4/6/97 & $B$ & 4.1 & 9.2 & 2.2 & &
\multicolumn{2}{c}{Low-frequency power present} \\*
       & 7/11/97 & $B$ & 3.6 & 7.8 & 2.2 & &
\multicolumn{2}{c}{Low-frequency power present}  \\*
       & 7/1/98 & $B$ & 2.5 & 5.5 & 2.2 & &
\multicolumn{2}{c}{Low-frequency power present} \\*
\\
\hline
\\
\multicolumn{9}{l}{$^a$Timing errors are larger than usual on this
date because the high-precision WWVB clock was broken.} \\
\multicolumn{9}{l}{$^b$Too little data to compute meaningful power spectrum.} \\
\end{tabular}
\end{table*}

\subsubsection{BX Mon}

\begin{figure}
\begin{center}
\epsfig{file=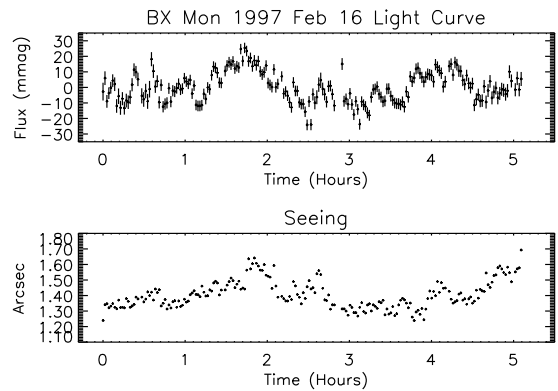,width=8cm}
\end{center}
\caption[1997 Feb 16 Light Curve and Seeing for BX Mon]{Light curve
and seeing measure for the 1997 Feb 16 observation of BX Mon.  Rapid
variability is detected, but its origin is questionable.  The
variations could either be intrinsic to BX Mon, or possibly due to
atmospheric changes that were imperfectly subtracted out by the
differential photometry. See text for details. \label{fig:bxmonfig1}}
\end{figure}

Our first observation of BX Mon, on 1997 Jan 18, produced only a small
amount of low-quality data, so we re-observed this object on 1997 Feb
16.  On that date, we found a significant excess variance.  The rms
fractional variation is 2.7 times higher than expected from Poisson
fluctuations and scintillation.  No comparison stars showed similar
variations, and the variations in BX Mon were evident no matter what
combination of comparison stars was used.  Unfortunately, there is some
chance that these variations are due to atmospheric changes.  A plot of
seeing as a function of time looks similar to the BX Mon light curve in
places, although the two quantities are not strictly correlated.  The 1997
Feb 16 light curve for BX Mon, and a measure of the PSF width (from a
different star in the same field) during the observation, are shown in
Figure \ref{fig:bxmonfig1}.  The seeing measure plotted in this figure is
the second moment of the distribution of counts within the aperture
used.  

\begin{figure}
\begin{center}
\epsfig{file=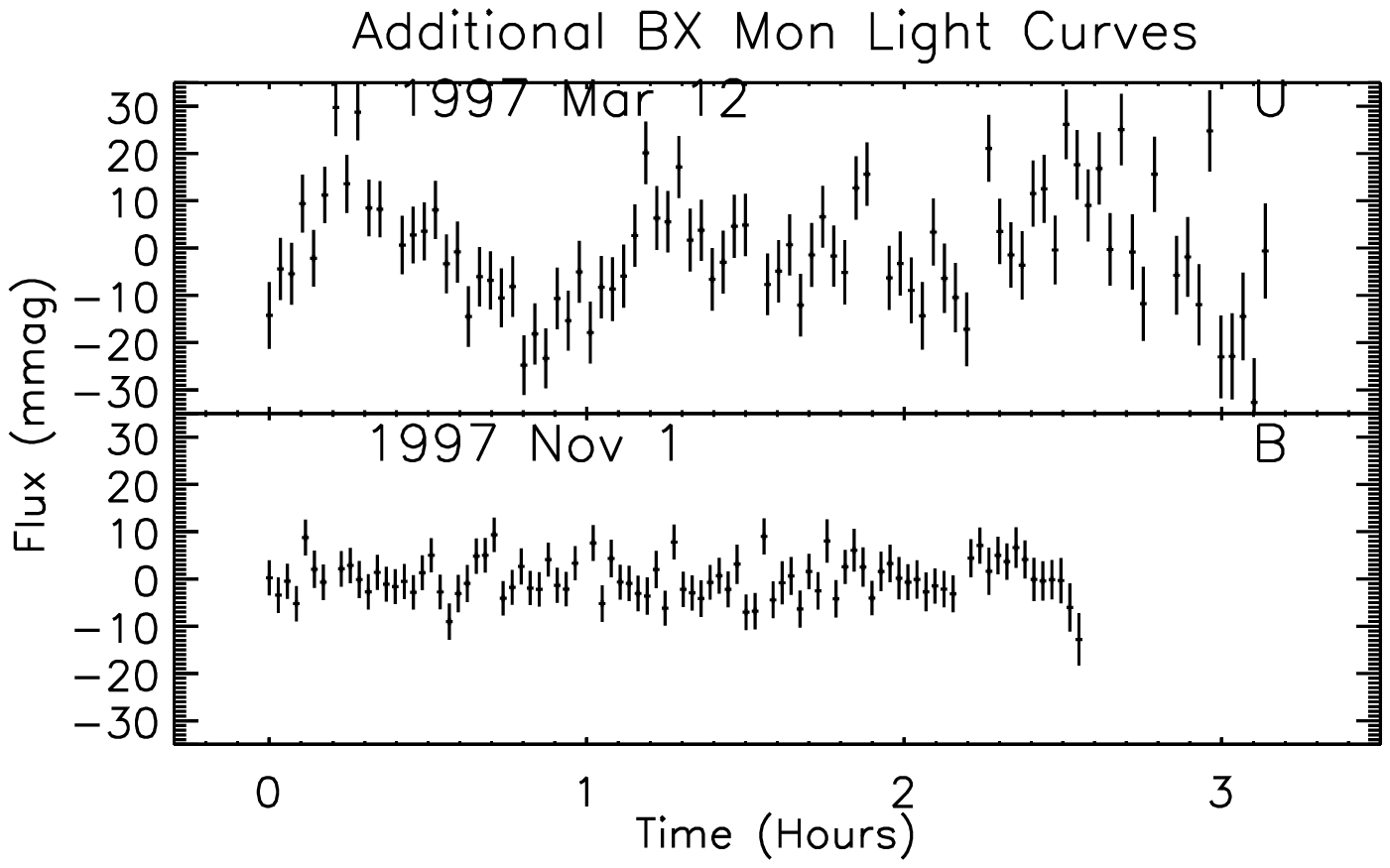,width=8cm}
\end{center}
\caption[Additional Light Curves for BX Mon]{Example light curves for
BX Mon.  Rapid variability is marginally detected in the $U$-band
observation on 1997 Mar 12 (top panel), but not in the $B$-band
observation on 1997 Nov 1 (bottom panel). \label{fig:bxmonfig2}}
\end{figure}

Follow-up observations were performed in 1997 Mar, Apr, and Nov, but
the results for this object remain inconclusive.  Flickering was
possibly detected in a $U$-band observation on 1997 Mar 12
($s/s_{exp}=1.9$), but on this night some of the comparison stars also
showed excess variability.  Variations were not detected in any other
observations.  Example light curves, from 1997 Mar 12 and 1997 Nov 1,
are shown in Figure
\ref{fig:bxmonfig2}.

\subsubsection{CM Aql}

\begin{figure}
\begin{center}
\epsfig{file=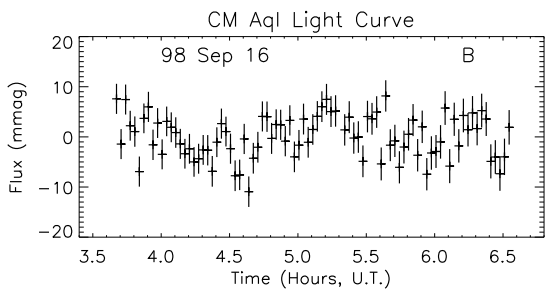,width=8cm}
\end{center}
\caption[CM Aql Light Curve]{$B$-band light curve for
CM Aql.  This observation is right at the border between a non-detection
and a marginal detection of rapid variability. \label{fig:cmaqllc}}
\end{figure}

We observed CM Aql once, on 1998 Sep 16, with a $B$ filter, using CCD
5. The measured fractional variation is just 35\% higher than expected
for this observation (i.e., $s/s_{exp}=1.35$).  However, the small
variations that are present (see Figure \ref{fig:cmaqllc}) do not
appear related to any of our house-keeping data, and the observation
was done in good weather conditions.  This situation is in contrast to
other borderline case observations with $s/s_{exp} \approx 1.4$ (e.g.,
TX CVn, CI Cyg, and AS 360), which were done under poor weather
conditions or do not show any point-to-point correlated variations.
Further observations are required to determine whether this flickering
is intrinsic to CM Aql.  It is possible that the higher-sensitivity
CCD 5 requires a more thorough study of systematic effects.

\subsubsection{BF Cyg}

\begin{figure}
\begin{center}
\epsfig{file=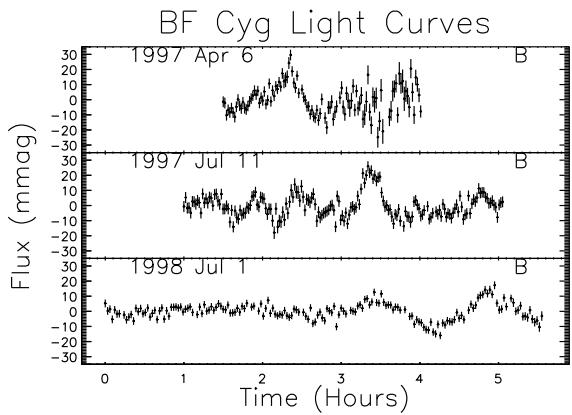,width=8cm}
\end{center}
\caption[BF Cyg Light Curves]{$B$-band light curves for
BF Cyg.  \label{fig:bfcyglcs}}
\end{figure}

We found low-amplitude variability in BF Cyg on three separate
occasions, separated by more than a year.  This variability does not
appear to be correlated with seeing or any other house-keeping data, and
it was not seen in any of the 5 comparison stars.  In each observation,
the measured variance was more than twice that expected from Poisson
variations and scintillation ($s/s_{exp} > 2$), and so this source is a
prime candidate for follow-up study.  All observations were done with
the $B$ filter.  The light curves are presented in Figure
\ref{fig:bfcyglcs}.

\subsection{Previously-Known Flickering Symbiotics}

\begin{figure}
\begin{center}
\epsfig{file=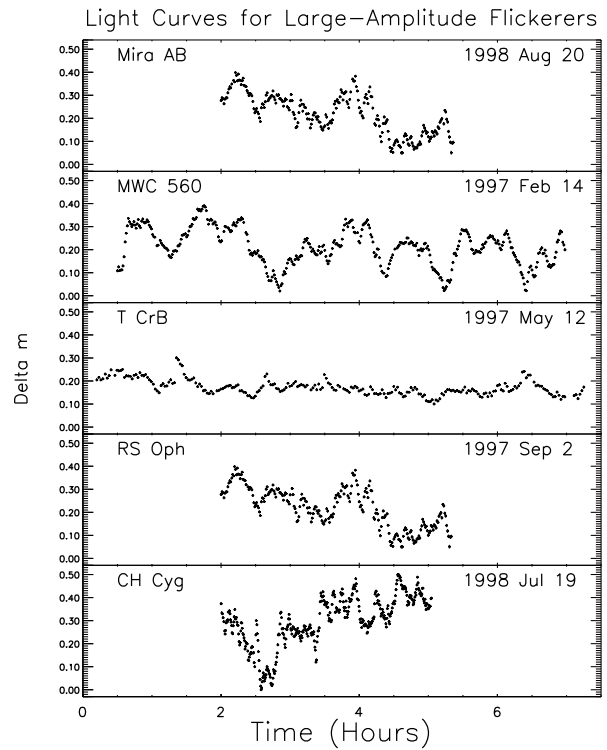,width=8.5cm}
\end{center}
\caption[Light Curves for the previously known flickering
symbiotics.]{Representative $B$-band light curves for the
previously-known flickering symbiotics. These objects form a sub-class
of symbiotics that are clearly distinct from the majority of systems,
in which rapid optical variability either has very low amplitude or is
undetectable.  \label{fig:flicklcs}}
\end{figure}

Included in our survey were five well-known flickering symbiotics:
Mira AB, MWC 560, T Crb, RS Oph, and CH Cyg.  We confirmed the
presence of rapid optical variability in all five systems, and
recorded a marked change in the flickering properties of both T CrB
and CH Cyg over the course of the survey.  A full analysis of these
observations is presented elsewhere (\nocite{sbh01}Sokoloski et
al. 2001b), but example light curves for each of the large-amplitude
flickerers are shown in Figure \ref{fig:flicklcs}.  These light curves
are shown, at least in part, to emphasize the difference between these
five systems and the other 30 objects in the sample.  Note that the
ordinate for each plot in Figure \ref{fig:flicklcs} spans 0.5
magnitudes, whereas the 25 objects for which no rapid variations were
detected are constant to within a few mmag.  The four new candidate
flickerers vary at a level of only a few tens of mmag.

\subsection{Future Work}

Although for the faintest targets a larger telescope would have
improved our sensitivity, in most cases we were more limited by the
capabilities of the CCD detectors than the telescope collecting area.
The limitations of the CCDs were especially apparent for the bright
objects, objects with bright comparison stars, or objects in sparse
fields.  When we observed fields containing bright stars, integration
times were limited by CCD non-linearity and saturation, and the duty
cycle of an observation was sometimes low.  To improve measurements of
low-amplitude rapid variability, one would like to have a CCD with a
larger field of view and faster readout times.  Masking techniques in
which part of the CCD is read out while another part is exposed would
greatly improve the efficiency of this type of observation.  A larger
field would allow for additional bright comparison stars.  For some
objects with no moderately bright stars nearby, this improvement is
absolutely essential for differential photometry.  For other fields,
having a larger number of comparison stars would allow for
point-spread-function fitting and a better understanding of
systematics.  In addition, the ideal detector would have either a
variable gain, or a gain that was high enough to allow for photometry
on a field of stars with a large dynamic range of brightnesses.  One
would like to always have the option of integrating to the
scintillation limit before saturating a star or reaching the
non-linear regime.  With such a detector, and observations done at low
airmass, one could probe much lower levels of variability ($< 1$
mmag), as well as shorter time scales.

\section{Discussion}\label{sec:disc}

Our survey has revealed new hints of rapid variability in four
symbiotics: EG And, CM Aql, BX Mon, and BF Cyg.  Our main result,
however, is that most symbiotics do not vary on a time scale of
minutes, down to a level of mmag.  Previous authors have noted this
apparent general lack of rapid variability in symbiotics
(e.g. \nocite{wal77}Walker 1977; \nocite{dob96}Dobrzycka et al. 1996).
Our work broadens and strengthens this conclusion.  We have placed
tight upper limits on both aperiodic and periodic variability for 25
of the 35 objects observed.  We also performed extensive observations
of the five previously-known large-amplitude flickerers, which are
found to be rare among symbiotics.  These observations, as well as
serendipitous discoveries of five new variable stars and results for
the magnetic symbiotic Z And, are presented in separate papers
(\nocite{sb99}Sokoloski \& Bildsten 1999; \nocite{sbcf01}Sokoloski et
al. 2001a; \nocite{sbh01}Sokoloski et al. 2001b).

This survey of optical variability has been carried out primarily in
the $B$-band, with some supplementary observations done in $U$. The
$B$-band optical flux from symbiotics originates from several sources,
including the red giant, recombination radiation from the nebula
(which is pumped by the far-ultraviolet and soft X-ray emission from
the hot component), and possibly some direct emission from a disk or
the hot star itself. For comparison, the optical flux in CVs is
dominated by emission from the accretion disk, and in supersoft
sources the optical light is primarily from an irradiated disk.  We
discuss below the differences in optical variability properties
between symbiotics and other accreting-WD binaries, differences among
the SS, and what these variability properties might be telling us
about the underlying sources of light.

One of the important distinctions between CVs and symbiotics is that
in many symbiotics, the high hot-component luminosity indicates that
material must be burning quasi-steadily on the surface of a white
dwarf, either because the accretion rate is in the steady-burning
regime, or as residual burning from a past outburst
(\nocite{pr80}Paczy{\'n}ski \& Rudak 1980; \nocite{iben82}Iben 1982;
\nocite{mk92}Miko{\l}ajewska \& Kenyon 1992; \nocite{siosta94}Sion \&
Starrfield 1994).  Some of the far-ultraviolet and soft X-ray flux
from this persistent nuclear burning is reprocessed into the optical
by the nebula.  Since the flux from the nuclear burning does not
change on short time scales, and the nebula itself is unlikely to
introduce rapid variability, the nuclear burning produces an extra
contribution to the optical flux in symbiotics that is not variable on
short time scales.  Therefore, the amplitude of any rapidly variations
from the accretion are reduced compared to similar variations in CVs.
In supersoft sources, as in symbiotics, far-ultraviolet and soft X-ray
emission from nuclear burning on a WD is reprocessed into the optical.
In supersoft sources, however, the reprocessing occurs primarily in a
flared accretion disk instead of an extended nebula
(\nocite{pds96}Popham \& DiStefano 1996), and
disk-rim fluctuations could introduce an additional source of rapid
optical variability (\nocite{mhsdbm98}Meyer-Hofmeister et al. 1998).

Thus the amplitude of rapid optical variability in symbiotics and
other accreting-WD binaries is linked to the presence or absence of a
physical process (nuclear burning) that profoundly affects the binary.
Nuclear burning on the surface of a WD produces much more power than
accretion alone, and the presence of such burning may be closely
related to the outbursts in classical symbiotics.  In one model for
these outbursts, the months-to-years-long events are due to the
expansion of the white dwarf photosphere in response to an increase in
the accretion rate onto the WD (see \nocite{mk92}Miko{\l}ajewska \&
Kenyon 1992 for a review of these models).  But this phenomenon, which
is similar to that proposed to explain the optical rises and X-ray
dips of the supersoft sources (e.g, \nocite{kah98}Kahabka 1998), can
only occur if the WD is burning material quasi-steadily.  And there is
another reason why it is critical to understand how the accumulated
fuel is burning. Symbiotics might be progenitors of Type Ia
supernovae, but this depends on their ability to increase in mass as
they accrete. The more fuel that is burned in steady-state, the better
for increasing the mass of the WD, since large classical nova
outbursts can actually excavate the WD.  The connection between the
amplitude of rapid optical variability and the presence of nuclear
burning on the WD is explored in greater detail in section \S
\ref{sec:nuc}.
  
\subsection{Magnetism in the Accreting White Dwarfs} 

In our survey of 35 symbiotics, we found one new clearly magnetic
system (Z And; see \nocite{sb99} Sokoloski \& Bildsten 1999).  For 25
objects, we detected no rapid variability, and we place upper limits
ranging from 1.9 to 13 mmag on periodic oscillation amplitudes in
these objects.  The absence of a detected periodic signal in these 25
systems, however, does not necessarily mean that they are not
magnetic.  The amplitude of the Z And oscillation during quiescence
was extremely small (roughly 1 mmag peak-to-peak), and the presence of
a similar-sized oscillation in the other objects is not ruled out by
any of the upper limits in Table \ref{tab:upperlims}.  In fact, the Z
And oscillation was only discovered because the object was
fortuitously observed during an outburst of the system, during which
the oscillation amplitude was larger.  It is worth noting here that
estimates of the luminosity of the WD in Z And gives values of 500 -
2000 $\lsun$ (\nocite{mur91}M{\"u}rset et al. 1991;
\nocite{dob96}Dobrzycka et al. 1996, and references therein),
indicating that nuclear burning must be occurring on the surface of the
WD.  Given our discussion above concerning the affect of nuclear
burning on the amplitude of rapid optical variations, it is not
surprising that the amplitude of the Z And oscillation is much
smaller than the corresponding optical modulations seen in
intermediate polars.

Searching for oscillations is more subtle in the flickering objects,
and we tackle this problem in a separate paper \nocite{sbh01}
(Sokoloski et al. 2001b). Therefore, here we only place a lower limit
on the fraction of symbiotics in our sample that contain magnetic
white dwarfs.  Since we detected one magnetic system out of 32
possible WD systems (three symbiotics in our sample have either
main-sequence-star or neutron-star hot components), at least 3\% of
the WD systems in our sample are strongly magnetic.  This number
should be compared to 2\% for field white dwarfs (53 out of 2249;
\nocite{ans99} Anselowitz et al. 1999), and 5 - 10\% for white dwarfs
in CVs (Warner 1995).  Although the statistics are obviously too small
to determine the true magnetic fraction for SS, our results are
consistent with this fraction being higher than it is in the field, as
with CVs.

\subsection{Flickering} 

Our survey did not yield any new strong ($s \simgt 20$ mmag)
flickerers.  For most objects, the upper limit we have placed on
aperiodic variability (flickering) is a few mmags (see Table
\ref{tab:upperlims}).  Weak flickering (at the $\simlt 10$ mmag level)
was possibly detected in a few of our survey objects, which we discuss
in \S \ref{sec:sresults}.  Unfortunately, flickering at this low level
is difficult to attribute to the source, as many observational
systematics can have a comparable amplitude.  The new candidate
flickerers are, however, clearly worth additional study.

By comparing the small fractional rms variations listed in Table
\ref{tab:upperlims} and Table \ref{tab:candvars} with the large flickering
amplitudes of the 5 well-known flickering symbiotics shown in Figure
\ref{fig:flicklcs}, it is apparent that the distribution of
flickering amplitudes in symbiotics is bimodal; there are a few
large-amplitude flickerers, but most symbiotics show little to no
flickering.  This is in sharp contrast to the CVs, where most show
persistent flickering at a level of $\simgt 100$ mmag.  The
flickering light in CVs could come from several areas, including the
inner accretion disk, the splash point, or the surface of the WD
\nocite{war95,bru92} (Warner 1995; Bruch 1992).  Moreover, these flickering sources
contribute to different degrees in different object.  But no matter
which specific origin, the power source is accretion, and flickering
is prevalent in the CVs. We have found that flickering is much less
prevalent in the symbiotics.  Several possible explanations include:
(a) the optical flux in most SS is dominated by light from the red
giant, which doesn't change on short time scales, (b) the nebula
``washes out" the flickering light, (c) most SS do not accrete though
a disk, and the wind-fed accretion does not produce optical
flickering, or (d) in SS with no, or only small-amplitude, rapid
variations, much of the hot component light is from nuclear burning of
accreted material, which is stable on a time scale of minutes.  We
consider each of these possibilities below.

\subsubsection{The Red Giant}

We can rule out this simple explanation by comparing the red giants in
the flickering and non-flickering symbiotics.  If the red giants in
the flickering symbiotics were systematically fainter, or of later
spectral type, than the red giants in the non-flickering symbiotics,
the contribution to the $B$-band flux from the red giant would be
lower.  This reduced contribution from the red giant could allow any
variable emission from the hot component that might normally be hidden
to be revealed.

In fact, there is no such systematic difference.  \nocite{kf87} Kenyon 
\& Fernandez-Castro (1987) derive
spectral types for the giants in 25 symbiotic binaries, and find most
systems to be M2-5 III.  Included in their study were three of the
well-known large-amplitude flickerers (T CrB, CH Cyg, and RS Oph), and
two of our marginal flickering detection objects (EG And and BX Mon).
The red giant in T CrB is found to be M$4.1\pm0.3$ III, that in CH Cyg
to be M$6.5\pm0.3$ III, and that in RS Oph to be K$5.7\pm0.4$ I-II.
The red giant in EG And was found to have spectral type M$2.4\pm0.3$
II, and the red giant in BX Mon was found to have type M$4.6\pm0.4$
III.  These giants are therefore all quite typical for symbiotics,
except for that in RS Oph, which is actually earlier and brighter than
average.  Furthermore, if the contribution from the red giant was the
main factor in determining the flickering amplitude, one would expect
a more continuous distribution of amplitudes.

\subsubsection{The Nebula}

Radio observations of thermal bremsstrahlung from the photo-ionized
nebulae in SS indicate that typical sizes are usually less than
$10^{15}$ cm \nocite{stb84} (Seaquist et al. 1984).  From UV line
intensity ratios, densities in the nebula are estimated to be $10^8$
to $10^{10}$ cm$^{-3}$, and temperatures are on the order of $10^4$ K
\nocite{ken86,nv87} (Kenyon 1986, Nussbaumer \& Vogel 1987).
The presence of the nebula could make flux modulation from either
magnetic accretion or flickering more difficult to detect in SS.

The nebula can affect the rapid optical variability in two ways.
Nebular reprocessing can hide some temporal variations from the hot
component which originates in the FUV or higher energies by acting as
a low-pass filter.  In addition, non-variable or slowly varying
optical emission from the nebula will reduce the amplitude of any
rapid variability that might originate in the optical band, for
example from an accretion disk.  Rapid variations in the flux of
ionizing photons could possibly be passed into the optical if the
nebula is dense and radiation bounded (i.e., if the ionized nebula is
in the form of a simple closed Str\"omgren region).  Roughly speaking,
if the nebula is radiation bounded, far-ultraviolet photons are all
absorbed in photo-ionization events.  If the output of ionizing
photons varies, then the size of the Str\"omgren region varies, and so
does the amount of reprocessed optical light.  If the nebula is dense
enough at the boundary between the ionized and neutral regions, the
recombination time will be short, and rapid variations in the ionizing
flux could in principle be propagated into the optical.  Typical
densities in the red giant winds that form the nebula can reach $\sim
10^{10}$ cm$^{-3}$, giving a hydrogen recombination time of minutes.
However, most SS nebulae are likely not radiation bounded (M\"urset et
al. 1991).  In addition, light travel time effects would cause any
sharp temporal features to be smeared out, as this light would be
reprocessed in parts of the nebula that are different distances away
from the hot white dwarf.  Finally, if the reprocessing time in the
nebula is long, then the rapid variability will not be passed along to
the optical light, no matter whether the nebula is radiation bounded
or density bounded (gas completely ionized so that some ionizing
photons escape).  The ionized nebula does not have as much effect on
optical photons, so variability which originates in the optical should
not be washed out, only possibly diluted.  Given the above discussion
about the difficulty of producing rapid variability from the nebular
emission, we would expect the spectrum of the rapidly-variable
component of the large-amplitude flickerers (such as RS Oph) to
reflect the physical origin of the variations, for example in an
accretion disk, and not be dominated by nebular features.

\subsubsection{Accretion Disk or Not?}
 
An interesting question to ask is whether the amplitude of flickering
in SS can tell us whether the accreted material's final plunge onto
the WD is via a disk or directly from the wind. The flickering in CVs
is related to disk accretion, and \nocite{liv88} Livio (1988) showed
that disks might well form in wind-fed symbiotics. A simple reason why
disk formation is possible is the large contrast between the accretion
radius, $r_a=GM/V^2 \approx 10^{13}$ cm (where $V$ is the relative
velocity between the red-giant wind and the WD), and the WD radius,
$R\approx 10^9$ cm.  In addition to being a known source of flickering
in CVs, a disk could naturally reveal much of this flickering in the
optical, and as we have mentioned above, optical photons are less
affected by the nebula then higher energy ionizing photons.
Unfortunately, we have no {\it a priori} phenomena to invoke to help
us know whether or not to expect optical flickering from direct
(i.e. no disk) wind accretors.  It is possible that material accreted
onto a white dwarf directly from an inhomogeneous wind could also
produce rapid variability, as is seen from wind accretion onto neutron
stars in X-ray binaries.  However, the emitted spectrum would not
necessarily have a large optical component, and variability from
direct, disk-less accretion might therefore be more likely to be
hidden.

Although the existence of accretion disks in symbiotics appears
likely, there is little direct evidence for disks around WDs in SS.
Disks are not needed in spectral fits for most systems \nocite{mur91}
(M\"urset et al. 1991) and double-peaked line profiles in most cases
cannot be definitively linked to disk emission
\nocite{rob94} (Robinson et al. 1994).  Observational evidence for a disk formed from a wind
has been found in Mira by \nocite{rc85} Reimers \& Cassatella (1985),
and \nocite{lp99} Lee \& Park (1999) have inferred the presence of an
accretion disk from double-peaked Raman scattering lines in RR Tel.
Sokoloski \& Bildsten (1999) have suggested that the small outburst of
Z And in 1997 is consistent with a disk-instability event, but the
inference of a disk there is indirect.

Finally, even if one did have direct evidence of a disk, the presence
of a disk does not guarantee the presence of a flickering source of
light.  Flickering is not seen in some dwarf novae near the maximum of
outburst
\nocite{war95} (Warner 1995), possibly due to the high $\mdot$.  As we have discussed,
symbiotics have higher accretion rates than most CVs, so it is
possible that the disks in SS would not flicker to the same extent
that CV disks do.  In summary, although there are reasons to believe
that accretion through a disk and disk-less wind accretion would
produce a different level of rapid optical variability, we are not yet
at the point of being able to use rapid optical variability as a
conclusive diagnostic of the form of accretion.

\subsubsection{Nuclear Power versus Gravitational Power, \\ and the
relationship between flickering and outbursts}

In their study of 8 SS, \nocite{dob96} Dobrzycka et al. (1996) note
that the objects with lower hot-component luminosities, $L_{hot}$,
seem more likely to flicker.  Our study of 35 objects produced 4 new
candidate small-amplitude flickering systems, but showed primarily
that most SS do not flicker (i.e., most have variations of less than a
few mmag).  Could this wide-spread lack of large-amplitude flickering
be due to the source of power in SS?  A potential hint to the
flickering puzzle can be found by considering the symbiotic recurrent
novae, which clearly burn most of their fuel in outbursts.  They are thus
unlikely to have large nuclear-burning luminosities in quiescence. The
expectation that these systems (between novae) are largely powered by
gravitational energy release is consistent with the detected
flickering.  For example, at least three of the four symbiotic
recurrent novae (RS Oph, T CrB, V1017 Sgr; we are not aware of
observations of the fourth, V3890 Sgr) flicker, and these three
systems also have $L_{hot} < 100
\,L_\odot$
\nocite{dob96} (Dobrzycka et al. 1996). Moreover, the flickering
amplitude in these systems is comparable to that seen in CVs
\nocite{wal77} (Walker 1977).  The other strong flickerers are MWC 560, CH Cyg, and
Mira.  Both Mira and CH Cyg have low hot component luminosities
($L_{hot} < 10\, L_\odot$).  In MWC 560, the hot component is
brighter, but the uncertainty in this quantity is large.  

Generally speaking, it is clear that the objects with low $L_{hot}$
preferentially exhibit CV-like flickering.  It is also the case that
the objects with higher $L_{hot}$ generally either do not flicker, or
have only small-amplitude flickering.  An interesting speculation that
follows from this observation is that any SS that are found to display
large-amplitude optical flickering should eventually experience a
nova.

A final bit of supporting evidence for the picture in which the
amplitude of the rapid optical variations is reduced in SS that
contain nuclear burning on a WD is the low amplitude of the coherent
oscillation in Z And compared to the DQ Her systems.  The optical
light of the DQ Hers typically is typically modulated by about 10\% at
the WD spin period
\nocite{war95,pat94} (Warner 1995; Patterson 1994), but in Z And the
modulation is only about $0.1$\% (when the system is in quiescence).
This modulation would be expected to be smaller than those seen in the
DQ Her systems if there is a large additional source of relatively
constant optical flux.  Note that this argument is based on the
assumption that the amount of pulsed optical flux from accretion onto
a magnetic WD is intrinsically the same for the two types of systems,
which may not in fact be the case.

\subsection{More on Nuclear Burning} \label{sec:nuc}

Most cataclysmic variables are clearly powered by gravitational energy
released in the process of accretion. In these low accretion-rate
systems ($\dot M<10^{-9} M_\odot \, {\rm yr^{-1}}$), the nuclear
burning of the accreted material occurs via classical novae and very
little fuel is burned between nova events. This should be contrasted
with the supersoft sources, where the high luminosity and spectral
fitting from ROSAT points to steady-state nuclear burning as the prime
power source for their emission at all times \nocite{vdh92} (van den
Heuvel et al. 1992). The required accretion rates for stable burning
range from $\dot M\approx 10^{-8} M_\odot\, {\rm yr^{-1}}$ for a
$M=0.4\, M_\odot$ WD to $\dot M\approx 3\times 10^{-7} M_\odot\, {\rm
yr^{-1}}$ for a $M=1.2\, M_\odot$ WD \nocite{pz78,fuj82} (Paczy\'{n}ski \& 
\.{Z}ytkow 1978; Fujimoto 1982).

The white dwarfs in symbiotics are thought to accrete at a rate of
$\sim 10^{-8} M_\odot \, {\rm yr^{-1}}$.  This rate is in the range
that can produce steady burning, and therefore forces us to consider
the importance of nuclear burning as an energy source in these
systems.  For accretion onto a white dwarf, nuclear energy release
exceeds that from gravity by a factor of $\sim 30$ (the ratio of the
accretion energy, $GM/R\approx 100-200 \ {\rm keV} \ nucleon^{-1}$, to
the nuclear burning energy, 5 MeV per nucleon, is 2-3\%), so that even
partial burning of the accreting fuel can prove important. Indeed,
$L_{hot}$ is so high for many symbiotics that gravitational energy
release cannot be the sole source of power, and burning must be
occurring. Careful spectral analysis of the nebular emission lines in
symbiotics is used to infer the luminosity, $L_{hot}$, of the hot
component, which is modeled as either a star or the inner region of an
accretion disk \nocite{mur91} (\nocite{kw84}Kenyon \& Webbink 1984;
M\"urset et al. 1991). We show these $L_{hot}$ estimates, where
available, in Table 3, along with other properties.  Furthermore, some
symbiotics have been detected with ROSAT as supersoft X-ray emitters
(\nocite{mur97}M\"urset et al. 1997).  It is not clear, however, that
this steady burning consumes all of the accreted fuel; possibly some
is still being accumulated and burns in an outburst.

A key point to note here is that the nuclear-burning luminosity would
not be expected to change on short time scales such as minutes or
hours.  Hence, a WD that is steadily burning the accreting material,
and reprocessing the burning emission into the optical, cannot flicker
rapidly at a high level.  The thermal time in a steadily burning
envelope at a depth denoted by the mass above it, $\Delta M$, is
$t_{th}\approx (C_PT/E_{nuc})(\Delta M/\dot M)$, where $C_P\approx
5k_B/2\mu m_p$ is the specific heat at constant pressure ($k_B$ is
Boltzmann's constant, and $\mu \approx 1/2$ is the mean molecular
weight), $T$ is the temperature in the envelope, and $E_{nuc}\approx
5\times 10^{18} \ {\rm erg \ g^{-1}}$ is the nuclear energy released
by hydrogen burning. For a typical temperature in a marginally stable
burning region of $T \approx 3\times 10^{7} {\rm K}$, the expression
above gives
\begin{equation} \label{eqn:nuctime}
t_{th}\approx 2\times 10^{-3}\left(\frac{\Delta M}{\dot M} \right). 
\end{equation}
For a $1.2 \ M_\odot$ WD, this time is approximately ten days, using
$\Delta M = 3 \times 10^{-6} M_\odot$ from Figure 7 in Fujimoto
(1982), and $\dot M = 2 \times 10^{-7} \ M_\odot \, {\rm
yr^{-1}}$. This time scale is shorter for more massive WDs, as $\Delta
M$ decreases with the mass of the WD. 
The shortest possible time is thus for a near-Chandraskhar-mass WD,
and even then equation (\ref{eqn:nuctime}) gives a time scale of about
one day (for $M = 1.4 \ M_\odot$, $\Delta M = 4 \times 10^{-7}
M_\odot$, and $\dot M = 3
\times 10^{-7} \ M_\odot \, {\rm yr^{-1}}$). So, it appears that the
thermal time at the location where the burning is occurring cannot get
shorter than about a day.

In symbiotics, a significant amount of the nuclear burning flux can
end up in the optical because of reprocessing in the nebula.  Since 
this flux does not vary rapidly, and the nebula is unlikely to
introduce rapid variations, any short-time-scale optical variations
due to accretion are like to be hidden or reduced, as we have found. 
In the supersoft sources, high-energy photons from the nuclear burning
can also be reprocessed into the optical, but in these systems it is
the large, flared disk that is the site of the reprocessing
\nocite{pds96} (Popham \& di Stefano 1996).  Photometric observations
with the required sensitivity are difficult to perform for most
supersoft sources, since they are generally fainter in the optical
than CVs or SS, but photometric studies have been done.  Rapid optical
variability was detected in the Galactic supersofts RX J0925.7-4758
\nocite{ccs01}(Clarkson et al. 2001), RX J0019.8+2156
(Meyer-Hofmeister et al. 1998), 1E 0035.4-7230 (Crampton et al. 1997;
van Teeseling et al. 1998), and possibly several others as well.
Meyer-Hofmeister et al. (1998) suggest that the variations in RX
J0019.8-2156 are due to changes in the disk rim, where the
nuclear-burning light is reprocessed.  The rapid variations in some of
the supersofts could also be due to CV-like flickering from the disk
that has not been completely hidden by the nuclear-burning light.  For
information on the effect of nuclear burning in CVs, we can look at
the short-period system T Pyx ($P_{orb} = 1.8$ hr).  T Pyx is believed
to be burning material to a significant extent even in quiescence
\nocite{pat98} (Patterson et al. 1998).  However, it does not have a
nebula or other large reprocessing site where the high-energy photons
can be converted to optical light, so the flux from the quasi-steady
burning in T Pyx probably makes only a relatively small contribution
to the optical flux.  In fact, the optical light curves from this
system show obvious large-amplitude flickering
\nocite{pat98} (Patterson et al. 1998), presumably from the accretion
disk.  So, in examining the relationship between the optical
variability properties of SS and other WD accretors, we find that the
presence or absence of nuclear burning on the WD can have an important
effect, but that the nature of the reprocessing site is also relevant.

\section*{Acknowledgments}

We would like to thank Mike Eracleous, James Graham, and Marten van
Kerkwijk for useful discussions, Mike Liu, Jonathan Baker, Greg
Ushomirsky, and Ed Brown for assistance with software, and the support
staff at Lick Observatory, especially Tony Misch, Rem Stone, and Will
Deitch.  We also thank the referee Krzysztof Belczy{\'n}ski for his
careful reading of the manuscript and useful comments. This research
was partially supported by NASA via grant NAG 5-8658 and by the
National Science Foundation under Grants PHY94-07194, PHY99-07949 and
AY97-31632.  L. B. is a Cottrell Scholar of the Research Corporation.

\bibliographystyle{mn}
\bibliography{mnrefs}  

\begin{thebibliography}{}

\bibitem[\protect\citeauthoryear{{Angel}, {Borra}, \& {Landstreet}}{{Angel}
  et~al.}{1981}]{abl81}
{Angel} J.~R.~P., {Borra} E.~F.,  {Landstreet} J.~D., 1981, \apjs, 45, 457

\bibitem[\protect\citeauthoryear{{Anselowitz} et~al.}{{Anselowitz}
  et~al.}{1999}]{ans99}
{Anselowitz} T., {Wasatonic} R., {Matthews} K., {Sion} E.~M.,  {McCook} G.~P.,
  1999, \pasp, 111, 702

\bibitem[\protect\citeauthoryear{{Bevington}}{{Bevington}}{1969}]{bev69}
{Bevington} P.~R., 1969, Data reduction and error analysis for the physical
  sciences.
\newblock New York: McGraw-Hill, 1969

\bibitem[\protect\citeauthoryear{{Bruch}}{{Bruch}}{1992}]{bru92}
{Bruch} A., 1992, \aap, 266, 237

\bibitem[\protect\citeauthoryear{{Bruch} \& {Duschl}}{{Bruch} \&
  {Duschl}}{1993}]{bd93}
{Bruch} A.,  {Duschl} W.~J., 1993, \aap, 275, 219

\bibitem[\protect\citeauthoryear{{Ciardullo} \& {Bond}}{{Ciardullo} \&
  {Bond}}{1996}]{cia96}
{Ciardullo} R.,  {Bond} H.~E., 1996, \aj, 111, 2332

\bibitem[\protect\citeauthoryear{{Clarkson}, {Charles}, \&
  {Sokoloski}}{{Clarkson} et~al.}{2001}]{ccs01}
{Clarkson} W., {Charles} P.~A.,  {Sokoloski} J.~L., 2001, in preparation

\bibitem[\protect\citeauthoryear{{Crampton} et~al.}{{Crampton}
  et~al.}{1997}]{chcs97}
{Crampton} D., {Hutchings} J.~B., {Cowley} A.~P.,  {Schmidtke} P.~C., 1997,
  \apj, 489, 903

\bibitem[\protect\citeauthoryear{{Dobrzycka}, {Kenyon}, \&
  {Milone}}{{Dobrzycka} et~al.}{1996}]{dob96}
{Dobrzycka} D., {Kenyon} S.~J.,  {Milone} A.~A.~E., 1996, \aj, 111, 414

\bibitem[\protect\citeauthoryear{{Downes} \& {Keyes}}{{Downes} \&
  {Keyes}}{1988}]{dk88}
{Downes} R.~A.,  {Keyes} C.~D., 1988, \aj, 96, 777

\bibitem[\protect\citeauthoryear{{Ezuka}, {Ishida}, \& {Makino}}{{Ezuka}
  et~al.}{1998}]{eim98}
{Ezuka} H., {Ishida} M.,  {Makino} F., 1998, \apj, 499, 388

\bibitem[\protect\citeauthoryear{{Fritz} \& {Bruch}}{{Fritz} \&
  {Bruch}}{1998}]{fb98}
{Fritz} T.,  {Bruch} A., 1998, \aap, 332, 586

\bibitem[\protect\citeauthoryear{{Fujimoto}}{{Fujimoto}}{1982}]{fuj82}
{Fujimoto} M.~Y., 1982, \apj, 257, 767

\bibitem[\protect\citeauthoryear{{Gilliland}}{{Gilliland}}{1992}]{gil92}
{Gilliland} R.~L., 1992, "Details of Noise Sources and Reduction Processes".
\newblock ASP Conference Series \ 23:Astronomical CCD Observing and Reduction
  Techniques, p.~68

\bibitem[\protect\citeauthoryear{{Gilliland} \& {Brown}}{{Gilliland} \&
  {Brown}}{1988}]{gb88}
{Gilliland} R.~L.,  {Brown} T.~M., 1988, \pasp, 100, 754

\bibitem[\protect\citeauthoryear{{Gilliland} \& {Brown}}{{Gilliland} \&
  {Brown}}{1992}]{gb92}
{Gilliland} R.~L.,  {Brown} T.~M., 1992, \pasp, 104, 582

\bibitem[\protect\citeauthoryear{{Groth}}{{Groth}}{1975}]{gro75}
{Groth} E.~J., 1975, \apjs, 29, 285

\bibitem[\protect\citeauthoryear{{Hayes} \& {Latham}}{{Hayes} \&
  {Latham}}{1975}]{hl75}
{Hayes} D.~S.,  {Latham} D.~W., 1975, \apj, 197, 593

\bibitem[\protect\citeauthoryear{{Hoard}}{{Hoard}}{1993}]{hoa93}
{Hoard} D.~W., 1993, \pasp, 105, 1232

\bibitem[\protect\citeauthoryear{{Howell}}{{Howell}}{1992}]{how92}
{Howell} S.~B., 1992, Introduction to Differential Time-Series Astronomical
  Photometry.
\newblock ASP Conference Series \ 23:Astronomical CCD Observing and Reduction
  Techniques, p. 105

\bibitem[\protect\citeauthoryear{{Iben}}{{Iben}}{1982}]{iben82}
{Iben} I., 1982, \apj, 259, 244

\bibitem[\protect\citeauthoryear{{Iben} \& {Tutukov}}{{Iben} \&
  {Tutukov}}{1996}]{it96}
{Iben} I.~J.,  {Tutukov} A.~V., 1996, \apjs, 105, 145

\bibitem[\protect\citeauthoryear{{Kahabka}}{{Kahabka}}{1998}]{kah98}
{Kahabka} P., 1998, \aap, 331, 328

\bibitem[\protect\citeauthoryear{{Kahabka} \& {van den Heuvel}}{{Kahabka} \&
  {van den Heuvel}}{1997}]{kvdh97}
{Kahabka} P.,  {van den Heuvel} E.~P.~J., 1997, \araa, 35, 69

\bibitem[\protect\citeauthoryear{{Kenyon}}{{Kenyon}}{1986}]{ken86}
{Kenyon} S.~J., 1986, "The symbiotic stars".
\newblock Cambridge and New York, Cambridge University Press, 1986, 295 p.

\bibitem[\protect\citeauthoryear{{Kenyon} \& {Fernandez-Castro}}{{Kenyon} \&
  {Fernandez-Castro}}{1987}]{kf87}
{Kenyon} S.~J.,  {Fernandez-Castro} T., 1987, \aj, 93, 938

\bibitem[\protect\citeauthoryear{{Kenyon} \& {Garcia}}{{Kenyon} \&
  {Garcia}}{1989}]{kg89}
{Kenyon} S.~J.,  {Garcia} M.~R., 1989, \aj, 97, 194

\bibitem[\protect\citeauthoryear{{Kenyon} et~al.}{{Kenyon}
  et~al.}{1991}]{kommsga91}
{Kenyon} S.~J., {Oliversen} N.~A., {Mikolajewska} J., {Mikolajewski} M.,
  {Stencel} R.~E., {Garcia} M.~R.,  {Anderson} C.~M., 1991, \aj, 101, 637

\bibitem[\protect\citeauthoryear{{Kenyon} \& {Webbink}}{{Kenyon} \&
  {Webbink}}{1984}]{kw84}
{Kenyon} S.~J.,  {Webbink} R.~F., 1984, \apj, 279, 252

\bibitem[\protect\citeauthoryear{{Leahy} et~al.}{{Leahy} et~al.}{1983}]{lea83}
{Leahy} D.~A., {Darbro} W., {Elsner} R.~F., {Weisskopf} M.~C., {Kahn} S.,
  {Sutherland} P.~G.,  {Grindlay} J.~E., 1983, \apj, 266, 160

\bibitem[\protect\citeauthoryear{{Lee} \& {Park}}{{Lee} \& {Park}}{1999}]{lp99}
{Lee} H.,  {Park} M., 1999, \apjl, 515, L89

\bibitem[\protect\citeauthoryear{{Livio}}{{Livio}}{1988}]{liv88}
{Livio} M., 1988, Accretion from Stellar Winds.
\newblock IAU Colloq.\ 103: The Symbiotic Phenomenon, p. 149

\bibitem[\protect\citeauthoryear{{Luthardt}}{{Luthardt}}{1992}]{lut92}
{Luthardt} R., 1992, in Reviews of Modern Astronomy, Vol.~5, p.~38

\bibitem[\protect\citeauthoryear{{Lyubarskii}}{{Lyubarskii}}{1997}]{lyu97}
{Lyubarskii} Y.~E., 1997, \mnras, 292, 679

\bibitem[\protect\citeauthoryear{{Meyer-Hofmeister} et~al.}{{Meyer-Hofmeister}
  et~al.}{1998}]{mhsdbm98}
{Meyer-Hofmeister} E., {Schandl} S., {Deufel} B., {Barwig} H.,  {Meyer} F.,
  1998, \aap, 331, 612

\bibitem[\protect\citeauthoryear{{Michalitsianos} et~al.}{{Michalitsianos}
  et~al.}{1993}]{mic93}
{Michalitsianos} A.~G. et~al., 1993, \apjl, 409, L53

\bibitem[\protect\citeauthoryear{{Mikolajewska} \& {Kenyon}}{{Mikolajewska} \&
  {Kenyon}}{992a}]{mk92axper}
{Mikolajewska} J.,  {Kenyon} S.~J., 1992a, \aj, 103, 579

\bibitem[\protect\citeauthoryear{{Mikolajewska} \& {Kenyon}}{{Mikolajewska} \&
  {Kenyon}}{992b}]{mk92}
{Mikolajewska} J.,  {Kenyon} S.~J., 1992b, \mnras, 256, 177

\bibitem[\protect\citeauthoryear{{Mikolajewski} \&
  {Mikolajewska}}{{Mikolajewski} \& {Mikolajewska}}{1988}]{mm88}
{Mikolajewski} M.,  {Mikolajewska} J., 1988, An Accretor-Propeller Model of CH
  Cygni.
\newblock IAU Colloq.\ 103: The Symbiotic Phenomenon, p. 233

\bibitem[\protect\citeauthoryear{{Mikolajewski}, {Mikolajewska}, \&
  {Khudiakova}}{{Mikolajewski} et~al.}{990b}]{mski90b}
{Mikolajewski} M., {Mikolajewska} J.,  {Khudiakova} T.~N., 1990b, \aap, 235,
  219

\bibitem[\protect\citeauthoryear{{Mikolajewski} et~al.}{{Mikolajewski}
  et~al.}{990a}]{mski90a}
{Mikolajewski} M., {Mikolajewska} J., {Tomov} T., {Kulesza} B.,  {Szczerba} R.,
  1990a, Acta Astronomica, 40, 129

\bibitem[\protect\citeauthoryear{{Muerset} et~al.}{{Muerset}
  et~al.}{1991}]{mur91}
{Muerset} U., {Nussbaumer} H., {Schmid} H.~M.,  {Vogel} M., 1991, \aap, 248,
  458

\bibitem[\protect\citeauthoryear{{Muerset}, {Wolff}, \& {Jordan}}{{Muerset}
  et~al.}{1997}]{mur97}
{Muerset} U., {Wolff} B.,  {Jordan} S., 1997, \aap, 319, 201

\bibitem[\protect\citeauthoryear{{Nussbaumer} \& {Vogel}}{{Nussbaumer} \&
  {Vogel}}{1987}]{nv87}
{Nussbaumer} H.,  {Vogel} M., 1987, \aap, 182, 51

\bibitem[\protect\citeauthoryear{{Paczynski} \& {Rudak}}{{Paczynski} \&
  {Rudak}}{1980}]{pr80}
{Paczynski} B.,  {Rudak} B., 1980, \aap, 82, 349

\bibitem[\protect\citeauthoryear{{Paczynski} \& {Zytkow}}{{Paczynski} \&
  {Zytkow}}{1978}]{pz78}
{Paczynski} B.,  {Zytkow} A.~N., 1978, \apj, 222, 604

\bibitem[\protect\citeauthoryear{{Patterson}}{{Patterson}}{1994}]{pat94}
{Patterson} J., 1994, \pasp, 106, 209

\bibitem[\protect\citeauthoryear{{Patterson} et~al.}{{Patterson}
  et~al.}{1998}]{pat98}
{Patterson} J. et~al., 1998, \pasp, 110, 380

\bibitem[\protect\citeauthoryear{{Popham} \& {di Stefano}}{{Popham} \& {di
  Stefano}}{1996}]{pds96}
{Popham} R.,  {di Stefano} R., 1996, in LNP Vol. 472: Supersoft X-Ray Sources,
  p.~65

\bibitem[\protect\citeauthoryear{{Reiger}}{{Reiger}}{1963}]{rei63}
{Reiger} S.~H., 1963, \aj, 68, 395

\bibitem[\protect\citeauthoryear{{Reimers} \& {Cassatella}}{{Reimers} \&
  {Cassatella}}{1985}]{rc85}
{Reimers} D.,  {Cassatella} A., 1985, \apj, 297, 275

\bibitem[\protect\citeauthoryear{{Robinson} et~al.}{{Robinson}
  et~al.}{1994}]{rob94}
{Robinson} K., {Bode} M.~F., {Skopal} A., {Ivison} R.~J.,  {Meaburn} J., 1994,
  \mnras, 269, 1

\bibitem[\protect\citeauthoryear{{Rodgers} et~al.}{{Rodgers}
  et~al.}{1997}]{rhbmor97}
{Rodgers} B., {Hoard} D.~W., {Burdullis} T., {Machado-Pelaez} L., {O'Toole} M.,
   {Reed} S., 1997, \pasp, 109, 1093

\bibitem[\protect\citeauthoryear{{Scargle}}{{Scargle}}{1982}]{sca82}
{Scargle} J.~D., 1982, \apj, 263, 835

\bibitem[\protect\citeauthoryear{{Seaquist} \& {Taylor}}{{Seaquist} \&
  {Taylor}}{1990}]{st90}
{Seaquist} E.~R.,  {Taylor} A.~R., 1990, \apj, 349, 313

\bibitem[\protect\citeauthoryear{{Seaquist}, {Taylor}, \& {Button}}{{Seaquist}
  et~al.}{1984}]{stb84}
{Seaquist} E.~R., {Taylor} A.~R.,  {Button} S., 1984, \apj, 284, 202

\bibitem[\protect\citeauthoryear{{Sion} et~al.}{{Sion} et~al.}{1988}]{sfml88}
{Sion} E.~M., {Fritz} M.~L., {McMullin} J.~P.,  {Lallo} M.~D., 1988, \aj, 96,
  251

\bibitem[\protect\citeauthoryear{{Sion} \& {Starrfield}}{{Sion} \&
  {Starrfield}}{1994}]{siosta94}
{Sion} E.~M.,  {Starrfield} S.~G., 1994, \apj, 421, 261

\bibitem[\protect\citeauthoryear{{Sokoloski} \& {Bildsten}}{{Sokoloski} \&
  {Bildsten}}{1999}]{sb99}
{Sokoloski} J.~L.,  {Bildsten} L., 1999, \apj, 517, 919

\bibitem[\protect\citeauthoryear{{Sokoloski} et~al.}{{Sokoloski}
  et~al.}{001a}]{sbcf01}
{Sokoloski} J.~L., {Bildsten} L., {Chornock} R.,  {Filippenko} A.~V., 2001a, in
  preparation

\bibitem[\protect\citeauthoryear{{Sokoloski}, {Bildsten}, \& {Ho}}{{Sokoloski}
  et~al.}{001b}]{sbh01}
{Sokoloski} J.~L., {Bildsten} L.,  {Ho} W., 2001b, in preparation

\bibitem[\protect\citeauthoryear{{Tomov} et~al.}{{Tomov} et~al.}{1992}]{tom92}
{Tomov} T., {Zamanov} R., {Kolev} D., {Georgiev} L., {Antov} A., {Mikolajewski}
  M.,  {Esipov} V., 1992, \mnras, 258, 23

\bibitem[\protect\citeauthoryear{{van den Heuvel} et~al.}{{van den Heuvel}
  et~al.}{1992}]{vdh92}
{van den Heuvel} E.~P.~J., {Bhattacharya} D., {Nomoto} K.,  {Rappaport} S.~A.,
  1992, \aap, 262, 97

\bibitem[\protect\citeauthoryear{{van der Klis}}{{van der Klis}}{1989}]{vdk89}
{van der Klis} M., 1989, "Fourier Techniques in X-Ray Timing".
\newblock Timing Neutron Stars, \ ed. H. Ogelman \& E. P. J. van den Heuvel \
  (Dordrecht: Kluwer Press), p.~27

\bibitem[\protect\citeauthoryear{{van Teeseling} et~al.}{{van Teeseling}
  et~al.}{1998}]{vtrpb98}
{van Teeseling} A., {Reinsch} K., {Pakull} M.~W.,  {Beuermann} K., 1998, \aap,
  338, 947

\bibitem[\protect\citeauthoryear{{van Zyl} et~al.}{{van Zyl}
  et~al.}{1999}]{vzyl99}
{van Zyl} L., {Warner} B., {O'Donoghue} D., {Sullivan} D., {Pritchard} J.,
  {Kemp} J., 1999, Baltic Astronomy, 9, 231

\bibitem[\protect\citeauthoryear{{Walker}}{{Walker}}{1977}]{wal77}
{Walker} A.~R., 1977, \mnras, 179, 587

\bibitem[\protect\citeauthoryear{{Warner}}{{Warner}}{1995}]{war95}
{Warner} B., 1995, "Cataclysmic variable stars".
\newblock Cambridge Astrophysics Series, Cambridge, New York: Cambridge
  University Press, |c1995

\bibitem[\protect\citeauthoryear{{Wickramasinghe} \&
  {Ferrario}}{{Wickramasinghe} \& {Ferrario}}{2000}]{wickfer00}
{Wickramasinghe} D.~T.,  {Ferrario} L., 2000, \pasp, 112, 873

\bibitem[\protect\citeauthoryear{{Willson}, {Garnavich}, \& {Mattei}}{{Willson}
  et~al.}{1981}]{wgm81}
{Willson} L.~A., {Garnavich} P.,  {Mattei} J.~A., 1981, Informational Bulletin
  on Variable Stars, 1961, 1

\bibitem[\protect\citeauthoryear{{Yonehara}, {Mineshige}, \&
  {Welsh}}{{Yonehara} et~al.}{1997}]{ymw97}
{Yonehara} A., {Mineshige} S.,  {Welsh} W.~F., 1997, \apj, 486, 388

\bibitem[\protect\citeauthoryear{{Young}}{{Young}}{1967}]{you67}
{Young} A.~T., 1967, \aj, 72, 747

\bibitem[\protect\citeauthoryear{{Yungelson} et~al.}{{Yungelson}
  et~al.}{1995}]{yun95}
{Yungelson} L., {Livio} M., {Tutukov} A.,  {Kenyon} S.~J., 1995, \apj, 447, 656

\bibitem[\protect\citeauthoryear{{Zamanov} \& {Bruch}}{{Zamanov} \&
  {Bruch}}{1998}]{zb98}
{Zamanov} R.~K.,  {Bruch} A., 1998, \aap, 338, 988

\end{thebibliography}

\appendix

\section[]{Average power of Poisson Noise with the Leahy normalization}

Using the \nocite{lea83}Leahy et al. (1983) normalization, we relate the average power in the power
spectrum to the rms variation of a light curve via
Parseval's theorem,
\begin{equation} \label{eqn:parseval}
\sum_{k=0}^{N-1} |c(k)|^2 = \frac{1}{N} \sum_{j=-N/2}^{N/2-1} |a_j|^2.
\end{equation}
The variance of c(k) is
\begin{equation}
\sigma^2 \equiv \frac{1}{N}\sum_{k=0}^{N-1}(c(k)-\overline{c})^2 =
\overline{c^2}-\overline{c}^2. 
\end{equation}
This variance should be the same as that given by equation
(\ref{eqn:sigmam2}) for a single star if observations are made under ideal
conditions, and airmass effects are removed from the data.  Plugging in
from equation (\ref{eqn:parseval}),
\begin{eqnarray}
\sigma^2 = \overline{c^2} - (\overline{c})^2 & = &
\frac{1}{N}\sum_{k=0}^{N-1}c(k)^2 - \left(\frac{1}{N}\sum_{k=0}^{N-1}
c(k) \right)^2 \\
 & = & \frac{1}{N^2}\sum_{j=-N/2}^{N/2-1} |a_j|^2 - \frac{1}{N^2}a_0^2\\
 & = &\frac{1}{N^2} \sum_{j=-N/2,j\neq 0}^{N/2-1} |a_j|^2 \\ 
 & = & \frac{1}{N^2} \sum_{j=-N/2,j\neq 0}^{N/2-1}\frac{C_{tot}P_j}{2} \\
 & = & \frac{C_{tot}}{N^2} \left( \sum_{j=1}^{N/2-1}P_j  + \frac{1}{2}P_{N/2} \right),
\end{eqnarray}
where in the last step we used the fact that $|a_j|=|a_{-j}|$.  The
fractional rms variation squared is then just
\begin{equation} \label{eqn:s2}
s^2 = \frac{\sigma^2}{\overline{c}^2} = \frac{N^2\sigma^2}{C_{tot}^2} =
\frac{1}{C_{tot}} \left( \sum_{j=1}^{N/2-1}P_j+ \frac{1}{2} P_{N/2}
\right).
\end{equation}
For a light curve with variations due to Poisson statistics only, we
expect $s^2 = 1/\overline{c} = N/C_{tot}$.  Setting the expression for
$s^2$ in equation (\ref{eqn:s2}) equal to our previous expression for
$s^2$, we find
\begin{equation}
\frac{N}{C_{tot}} = \frac{1}{C_{tot}} \left( \sum_{j=1}^{N/2-1}P_j+ \frac{1}{2} P_{N/2}
\right),
\end{equation}
or
\begin{equation}
\frac{2}{N} \left( \sum_{j=1}^{N/2-1}P_j+ \frac{1}{2} P_{N/2} \right) = 2
\end{equation}
\begin{equation}
\Rightarrow \overline{P} = 2 + \underbrace{\frac{P_{N/2}}{N}}_{\rm
small} \approx 2,
\end{equation}
where $\overline{P}$ is the average power excluding the D.C. power $P_0$.
So, for the power spectrum of Poisson noise with this normalization, the
mean value of the power at non-zero frequencies is approximately 2.

\bsp

\label{lastpage}

\end{document}